\documentclass[12pt,letter]{article}
\pdfoutput=1
\usepackage{graphicx, epsfig, color,cite}
\usepackage{amsmath}
\usepackage{amssymb}
\usepackage{float}
\usepackage{caption,subcaption,graphicx}
\usepackage{hyperref}

\def\eslt{\not\!\!\!{E_T}}
\def\to{\rightarrow}

\def\bi{\begin{itemize}}
\def\ei{\end{itemize}}

\def\tchi{\tilde\chi}

\def\tst{\tilde t}

\def\tg{\tilde g}

\def\alt{\lesssim}
\def\agt{\gtrsim}
\def\be{\begin{equation}}  
\def\ee{\end{equation}}  
\def\bea{\begin{eqnarray}}  
\def\eea{\end{eqnarray}}

\begin{document}
\begin{titlepage}
\begin{flushright}
OU-HEP-230601
\end{flushright}

\vspace{0.5cm}
\begin{center}
  {\Large \bf Prospects for charged Higgs bosons in natural \\
     SUSY models at the high-luminosity LHC
    }\\
\vspace{1.2cm} \renewcommand{\thefootnote}{\fnsymbol{footnote}}
{\large Howard Baer$^{1,2}$\footnote[1]{Email: baer@ou.edu },
Vernon Barger$^2$\footnote[2]{Email: barger@pheno.wisc.edu},
Xerxes Tata$^3$\footnote[3]{Email: tata@phys.hawaii.edu} and
Kairui Zhang$^2$\footnote[3]{Email: kzhang89@wisc.edu}
}\\ 
\vspace{1.2cm} \renewcommand{\thefootnote}{\arabic{footnote}}
{\it 
$^1$Homer L. Dodge Department of Physics and Astronomy,
University of Oklahoma, Norman, OK 73019, USA \\[3pt]
}
{\it 
$^2$Department of Physics,
University of Wisconsin, Madison, WI 53706 USA \\[3pt]
}
{\it 
$^3$Department of Physics and Astronomy,
University of Hawaii, Honolulu, HI 53706 USA \\[3pt]
}

\end{center}

\vspace{0.5cm}
\begin{abstract}
\noindent
We continue our examination of prospects for discovery of heavy
Higgs bosons of natural SUSY (natSUSY) models at the high luminosity LHC
(HL-LHC), this time focussing on charged Higgs bosons.  
In natSUSY, higgsinos are expected at the few
hundred GeV scale whilst electroweak gauginos inhabit the TeV scale and
the heavy Higgs bosons, $H$, $A$ and $H^\pm$ could range up tens of TeV
without jeopardizing naturalness.
For TeV-scale heavy SUSY Higgs bosons $H$, $A$ and $H^\pm$, as currently
required by LHC searches, SUSY decays 
into gaugino plus higgsino  can dominate $H^\pm$ decays
provided these decays are kinematically accessible.
The visible decay products of higgsinos are soft making them
largely invisible, whilst the gauginos decay to $W$, $Z$ or $h$ plus
missing transverse energy ($\eslt$).  Charged Higgs bosons are
dominantly produced at LHC14 via the parton subprocess, $gb\to H^\pm t$.
In this paper, we examine the viability of observing signtures from
$H^\pm \to \tau\nu$, $H^\pm\to tb$ and $H^\pm \to W, Z, h + \eslt$
events produced in association  with a top quark at the
HL-LHC over
large Standard Model (SM) backgrounds from (mainly) $t\bar{t}$,
$t\bar{t}V$ and $t\bar{t}h$ production (where $V=W,\ Z$).  We find
that the greatest reach is found via the SM $H^\pm(\to\tau\nu) +t$
channel with a subdominant contribution from the $H^\pm(\to tb) +t$
channel.  Unlike for neutral Higgs searches, the SUSY decay modes
appear to be unimportant for $H^\pm$ searches at the HL-LHC.
We delineate regions of the $m_A$ vs. $\tan\beta$ plane, mostly
around $m_A \sim 1-2$~TeV, where signals from charged Higgs bosons would
serve to confirm signals of a heavy, neutral Higgs boson at the
$5\sigma$ level or, alternatively, to exclude heavy Higgs bosons at the
95\% confidence level at the high luminosity LHC.
\end{abstract}
\end{titlepage}

\section{Introduction}
\label{sec:intro}

There are two routes to discovery of supersymmetry (SUSY) at hadron
colliders such as the CERN Large Hadron Collider (LHC): one is via
direct pair production of $R$-parity odd states and the other is via
single (or pair) production of new $R$-parity even states such as the
additional heavy Higgs bosons present in the Minimal Supersymmetric
Standard Model (MSSM)\cite{Baer:2006rs}. Of course, strictly speaking,
the production of the heavy Higgs boson states is not necessarily a
signal for supersymmetry (unless these are seen via their decays into
supersymmetric particles) since such states are also possible in
non-supersymmetric models with an extended Higgs sector.  In the
present paper, we continue our work on the second approach: prospects
for SUSY discovery via the required additional SUSY Higgs bosons.
While much work has been done in this field, our focus is on LHC
signals of the heavy Higgs bosons of {\it natural} SUSY models,
wherein no large fine-tunings are required in order to gain a weak
scale characterized by $m_{W,Z,h}\simeq 100$ GeV.  In previous work,
we examined prospects for SUSY Higgs discovery in natural SUSY via
resonance production of heavy neutral Higgs bosons $H$ and $A$,
followed by 1. decays into Standard Models modes with
$H,\ A\to\tau^+\tau^-$ being most promising\cite{Baer:2022qqr}, and
2. decays into pairs of SUSY particles\cite{Baer:2022smj} which offer
qualitatively new channels for SUSY Higgs boson discovery.  Within
natural SUSY, once the heavy Higgs boson decay channels to
gaugino+higgsino become open, these may rapidly dominate the branching
fractions. This leads to two effects: 1. the new SUSY decay modes
diminish the branching fractions into SM modes, thus diminishing the
expected LHC reach via these  decay channels, and 2. the new
SUSY decay modes open up new avenues for SUSY Higgs detection, where
these new channels would signal the presence of the expected SUSY
particles.  In the present paper, we extend our earlier analyses to include
production of charged SUSY Higgs bosons $H^\pm$.  Discovery of charged
SUSY Higgs bosons is expected to be more challenging than discovery of
the neutral bosons. This is due to typically smaller production cross
sections (for a given Higgs boson mass) but also to less distinctive
discovery signatures. We investigate here whether this situation still
maintains under the rubric of natural SUSY.

Here, we take the measured value of the $Z$-boson mass as
representative of the magnitude of weak scale,
where in the MSSM the $Z$ mass is related to the weak scale
Lagrangian parameters via the electroweak minimization condition
\be
m_Z^2/2 =\frac{m_{H_d}^2+\Sigma_d^d-(m_{H_u}^2+\Sigma_u^u )\tan^2\beta}{\tan^2\beta -1}-\mu^2
\label{eq:mzs}
\ee where $m_{H_u}^2$ and $m_{H_d}^2$ are the Higgs soft breaking
masses, $\mu$ is the (SUSY preserving) superpotential higgsino mass parameter
and the $\Sigma_d^d$ and $\Sigma_u^u$ terms contain a large assortment
of loop corrections (see Appendices of Ref. \cite{Baer:2012cf} and
\cite{Baer:2021tta} and also \cite{Dedes:2002dy} for leading two-loop
corrections).  We adopt the notion of {\it practical
  naturalness}\cite{Baer:2015rja,Baer:2023cvi}, wherein the value of an observable
${\cal O}$ is natural if all {\it independent} contributions to 
${\cal  O}$ are comparable to (within a factor of few), or smaller than ${\cal
  O}$.  For natural SUSY models, we use the naturalness
measure\cite{Baer:2012up,Baer:2012cf} \be \Delta_{EW}\equiv |{\rm
  maximal\ term\ on\ the\ right-hand-side\ of\ Eq.~(\ref{eq:mzs})}|/(m_Z^2/2)\;,
\ee and take \be \Delta_{EW}\alt 30
\label{eq:dew30}
\ee
to be natural.
For most SUSY benchmark models, the superpotential $\mu$ parameter is tuned
to cancel against large contributions to the weak scale from SUSY breaking.
Since the $\mu$ parameter typically arises from very different physics
than SUSY breaking, {\it e.g.} from whatever solution to the SUSY
$\mu$ problem that is assumed,\footnote{Twenty solutions to the SUSY
  $\mu$ problem are recently reviewed in Ref. \cite{Bae:2019dgg}.}
then such a ``just-so'' cancellation is highly implausible\cite{Baer:2022dfc}
(though logically possible) compared to the case where all contributions
to the weak scale are $\sim m_{weak}$,
so that $\mu$ (or other parameters) need not be tuned.

Several important implications of Eq. (\ref{eq:dew30}) for
heavy SUSY Higgs searches include the following.
\begin{itemize}
\item The superpotential $\mu$ parameter enters $\Delta_{EW}$ directly,
  leading to $|\mu |\alt 350$ GeV.
  This implies that for heavy Higgs searches with $m_{H^\pm}\agt 2|\mu |$, then
  SUSY decay modes of $H^\pm$ should typically be open.
  If these additional
  decay widths to SUSY particles are large, then the branching fractions to
  the (usually assumed) SM search modes would be correspondingly
  reduced.

\item For $|m_{H_d}^2|\gg |m_{H_u}^2|$ or $\mu^2$, then $|m_{H_d}|$ sets the
  heavy Higgs mass scale ($m_{A,H,H^\pm}\sim |m_{H_d}|$) while
  $|m_{H_u}|$ sets the mass scale for $m_{W,Z,h}$.  Then, assuming that
  $m_{H^\pm} \gg M_W$, naturalness requires\cite{Bae:2014fsa}
    \end{itemize}
\be
m_{A,H,H^\pm}\alt m_Z\tan\beta\sqrt{\Delta_{EW}}.
\ee
For $\tan\beta\sim 10$ with $\Delta_{EW}\alt 30$, then $m_{H^\pm}$ can range up
to $\sim 5$ TeV. For $\tan\beta\sim 40$, then $m_{H^\pm}$ stays natural up to
$\sim 20$ TeV (although for large $\tan\beta\agt 20$, then bottom squark
contributions to $\Sigma_u^u$ become large and provide much
stronger upper limits on natural SUSY spectra\cite{Baer:2015rja}).

In Sec. \ref{sec:BM}, we first present a natural SUSY benchmark point
which then leads to a natural SUSY Higgs scenario which we previously
dubbed $m_H^{125}({\rm nat})$.  The $m_h^{125}({\rm nat})$ scenario is
promoted as a template for SUSY Higgs searches in that\cite{Baer:2022qqr}
1. it leads to a value of $m_h\simeq 125$ GeV throughout almost the entire
$m_A$ vs. $\tan\beta$ search plane and
2. the value of $\Delta_{EW}$ is also
low (though sometimes exceeding a value of 30 at higher $\tan\beta$
values) thoughout the search plane.  In Sec. \ref{sec:prod}, we list
the dominant charged SUSY Higgs boson production cross sections at
LHC14 (LHC with $\sqrt{s}=14$ TeV).
It is well-known that the $gb\to tH^\pm$ subprocesses is the dominant
$H^\pm$ production mechanism at the
LHC\cite{Bawa:1989pc,Gunion:1993sv,Barger:1993th}.
In Sec. \ref{sec:decay}, we present charged Higgs boson branching
fractions from the $m_h^{125}({\rm nat})$ scenario in the $m_A$ vs.
$\tan\beta$ plane, and find that indeed SUSY decay modes do become
rapidly dominant once these are kinematic accessible.  In
Sec. \ref{sec:taunu}, we examine $pp\to tH^\pm + X$ followed by
$H^\pm\to \tau^\pm\nu_{\tau}$ and map out distributions which help
obtain signal over SM background levels.  In Sec. \ref{sec:tb}, we
examine $pp\to tH^\pm + X$ followed by $H^\pm\to tb$ decay. In
Sec.~\ref{sec:HCsusy}, we study the impact of the SUSY decays of
$H^\pm$ bosons: unfortunately, we find that the relevant
cross-sections are mostly in the sub-fb range for mass values where
the branching fractions for these modes become substantial, and
(contrary to what we found for $H$ and $A$ bosons) SUSY decays of
$H^\pm$ do not offer a viable search strategy.  In
Sec. \ref{sec:reach}, we plot out the reach of high-luminosity LHC for
charged Higgs bosons of natural SUSY.  Our summary and conclusions are
contained in Sec. \ref{sec:conclude}.

\subsection{A review of some previous related work}

Here, we present a brief (and likely incomplete) review of some related
work on charged Higgs bosons from SUSY.  
In Ref. \cite{Baer:1985hd}, it was already emphasized that
detection of a top quark signal in accord with SM expectations would preclude
the decay $t\to bH^+$ and thus require $m_{H^\pm}\agt m_t-m_b$.
In light of present $t\bar{t}$ signal results,
this implies $m_{H^\pm}\agt 168$ GeV.
This result was already used by Kunszt and Zwirner in Ref. \cite{Kunszt:1991qe}
to form the low $m_A$ limit of the proposed $m_A$ vs. $\tan\beta$
heavy SUSY Higgs search plane.
Decays of heavy SUSY Higgs boson to SUSY decay modes were originally explored
in Ref. \cite{Baer:1987eb,Gunion:1987ki,Gunion:1988yc,Baer:1992kd}
and the complete set of $H^\pm $ decay widths may be found in Appendix C of
\cite{Baer:2006rs}.
In Ref. \cite{Bae:2014fsa}, these decay modes were examined
in the context of natural SUSY.
In that work, it was noted that for $H,\ A,\ H^\pm \to wino+higgsino$ channels,
the higgsino decays led to mainly soft, quasi-visible
decay debris whilst the winos decayed dominantly via two-body modes into
$W+higgsino$, $Z+higgsino$ and $h+higgsinos$.

In Ref. \cite{Bawa:1989pc,Gunion:1993sv,Barger:1993th}, it was found
that the dominant production process for charged Higgs bosons at LHC was
the reaction $gb\to tH^-+c.c.$.  NLO corrections to this production
process were calculated in Ref. \cite{Plehn:2002vy,Berger:2003sm} and
\cite{Zhu:2001nt}.  Signals from the final state $tH^\pm(\to
\tau\nu_{\tau})$ were examined in Ref. \cite{Roy:1999xw} and
\cite{Assamagan:2002in}.  The decay channel $H^+\to hW^+$ (which is
highly suppressed in natSUSY) was examined in Ref. \cite{Drees:1999sb}.
The use of three\cite{Moretti:1999bw} and four\cite{Miller:1999bm}
$b$-quark tags in $gg\to tbH^\pm$ was examined.  Corrections to the
$tbH^\pm$ vertex were examined in Ref. \cite{Carena:1999py}.  Initial
projections of the LHC reach for (charged) SUSY Higgs bosons were given
by Denegri {\it et al.}\cite{Denegri:2001pn} and by Assamagan {\it et
  al.}\cite{Assamagan:2002ne}.  Search limits from LHC for the $H^\pm\to
\tau\nu_{\tau}$ decay mode were presented based on $\sim 36$ fb$^{-1}$
of integrated luminosity by ATLAS\cite{ATLAS:2018gfm} and by
CMS\cite{CMS:2019bfg}.  An ATLAS search for charged Higgs bosons in the
$H^\pm tb$ mode based on 139 fb$^{-1}$ was given in
Ref. \cite{ATLAS:2021upq}.
A guidebook for LHC searches for SUSY and non-SUSY charged
Higgs bosons was provided in Ref. \cite{Arhrib:2018ewj} and a
review of non-SUSY charged Higgs is available in Ref. \cite{Akeroyd:2016ymd}.

\section{A natural SUSY benchmark point and the $m_h^{125}({\rm nat})$ scenario}
\label{sec:BM}

Following our previous work, we here adopt the same natural SUSY benchmark
point as in Ref. \cite{Baer:2022smj}, which was dubbed
$m_h^{125}({\rm nat})$ since the value of $m_h$ is very close to its
measured value throughout the entire $m_A$ vs. $\tan\beta$ plane.  We use the
two-extra-parameter non-universal Higgs model (NUHM2)\cite{Baer:2005bu}
with parameter space $m_0,\ m_{1/2},\ A_0,\ \tan\beta,\ \mu,\ m_A$ which
is convenient for naturalness studies since $\mu$ can be set to its
natural range of $\mu\sim 100-350$ GeV whilst both $m_A$ and $\tan\beta$
are free parameters.\footnote{The NUHM2 framework allows for independent
  soft SUSY breaking mass parameters for the scalar fields $H_u$ and
  $H_d$ in the Higgs sector, but leaves the matter scalar mass
  parameters universal to avoid flavour problems. The parameters
  $m_{H_u}^2$ and $m_{H_d}^2$ are then traded for $\mu$ and $m_A$ in the
  parameter set shown in (\ref{eq:param}).} We adopt the following natural SUSY
benchmark Higgs search scenario: \be m_h^{125}({\rm nat}):\ m_0=5\ {\rm
  TeV},\ m_{1/2}=1\ {\rm TeV},\ A_0=-1.6m_0,\ \tan\beta
,\ \mu=200\ {\rm GeV}\ {\rm and}\ m_A . \label{eq:param} \ee 

A similar $m_h^{125}({\rm nat})$ benchmark model spectrum, but with $\mu
=250$ GeV and $m_{1/2}=1.2$ TeV, was shown in Table 1 of
Ref. \cite{Baer:2022qqr} for $\tan\beta =10$ and $m_A=2$ TeV and so for
brevity we do not show the revised spectrum here.  We adopt the computer
code Isajet\cite{Paige:2003mg} featuring Isasugra\cite{Baer:1994nc} for
spectrum generation.  The SUSY Higgs boson masses are computed using
renormalization-group (RG) improved third generation fermion/sfermion
loop corrections\cite{Bisset:1995dc}.  The RG improved Yukawa couplings
include full threshold corrections\cite{Pierce:1996zz} which account for
leading two-loop effects\cite{Carena:2002es}.  For $\tan\beta =10$ and
$m_A=2$ TeV, we note that $\Delta_{EW}=16$ so the model is indeed EW
natural.  Also, with $m_h=124.6$ GeV, $m_{\tg}=2.4$ TeV and
$m_{\tst_1}=1.6$ TeV, it is consistent with LHC Run 2 SUSY search
constraints.  Most important for our purpose, the two lightest
neutralinos, $\tchi_1^0$ and $\tchi_2^0$, and the lighter chargino,
$\tchi_1^\pm$, are higgsino-like with masses $\sim 200$ GeV while the
neutralino $\tchi_3^0$ is bino-like with a mass of 450~GeV and the
heaviest neutralino and the heavier chargino have masses $\sim
0.86$~TeV.  Thus, the $H,\ A,\ H^\pm\to wino+higgsino$ decay modes turn
on for $m_{H,A,H^\pm}\agt 1.1$ TeV (although $H,\ A, \ H^\pm \to
bino+higgsino$ turns on at somewhat lower $m_A$ values). It is important
to note that while the value of $\Delta_{EW}$ may change somewhat for
small variations of the parameters that are held fixed in
Eq.~(\ref{eq:param}), we expect that the Higgs sector phenomenology  is
relatively insensitive to our specific choice
(as long as $\mu \alt 300$~GeV to maintain naturalness). 

\section{$H^\pm$ production cross sections at LHC14}
\label{sec:prod}

\begin{figure}[htb!]
\begin{center}
\includegraphics[height=0.5\textheight]{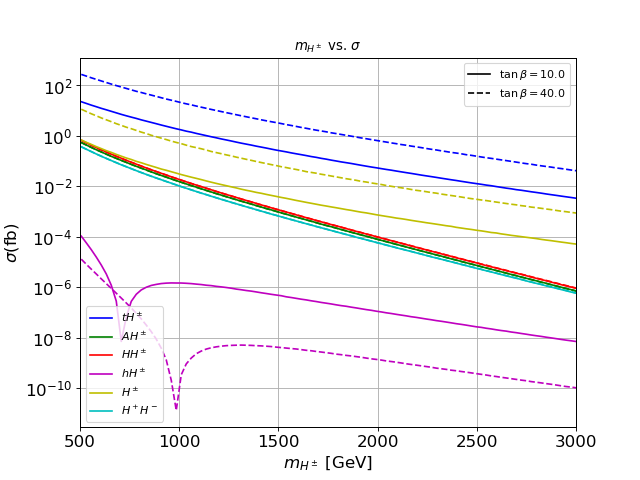}
\caption{The total cross section for $pp\to H^\pm +X$
via various production mechanisms at LHC14 for $\tan\beta =10$ (solid)
and $\tan\beta =40$ (dashed). 
\label{fig:sig1}}
\end{center}
\end{figure}

In Fig. \ref{fig:sig1}, we show leading order total cross sections for
production of charged SUSY Higgs bosons at LHC14, as generated using
Pythia\cite{Sjostrand:2007gs} for two values of $\tan\beta =10$ (solid)
and 40 (dashed).
We show cross sections for charged Higgs boson production at the LHC
via $tH^\pm$ production (blue), resonant $H^\pm$ production (light green),
$H^+H^-$ production (light blue), and $hH^\pm$ (magenta), $AH^\pm$ (green)
and $HH^\pm$ (red) production.

We see from Fig. \ref{fig:sig1} that the dominant production mechanism
at LHC14 is via the $gb\to tH^\pm$ subprocess
\cite{Bawa:1989pc,Gunion:1993sv,Barger:1993th}.  For $\tan\beta=40$, the
cross section varies from $\sim 10^3$ fb at low $m_{H^\pm}$ to $\sim
10^{-2}$ fb at $m_{H^\pm}\sim 3$ TeV.  These dominant cross sections are
enhanced by large $\tan\beta$.  If we compare $tH^\pm$ production to the
rate for $b\bar{b},gg\to A,\ H$ production\cite{Baer:2022qqr} for
$\tan\beta =10$ and $m_A\sim 1.5$ TeV, we find that charged Higgs
production rates are suppressed compared to resonance production of $A$
by a factor $\sim 5$. And since $\sigma (b\bar{b}, gg\to A)\sim \sigma
(b\bar{b},gg\to H)$, charged Higgs production compared to resonance
$H,\ A$ production is suppressed by about an order of magnitude. This
seems reasonable since charged Higgs production occurs in association
with a (spectator) top quark in contrast to the resonantly produced
neutral $H$ or $A$ boson.
%

The next largest cross section is direct resonance production of $H^\pm$
via (Yukawa suppressed) $u\bar{d}$ fusion or via (parton distribution
function suppressed) $c\bar{s}$ fusion. Note that this is suppressed
relative to neutral $H/A$ resonance production from $b\bar{b}$ fusion by
the smaller Yukawa couplings (of the first two generation of quarks)
and/or small Kobayashi Maskawa mixing elements.  These light green curves show
a $\tan\beta$ enhancement due to the Yukawa couplings involved in the
production mechanism. The resonance production cross section is
typically about two orders of magnitude below $tH^\pm$ associated
production.

The next largest production cross sections are
$HH^\pm$ and $AH^\pm$ which are produced via $W^*$ exchange
followed closely by $H^+H^-$ pair production
which takes place via $s$-channel $\gamma^*$ and $Z^*$ exchange.
All these production vertices involve gauge interactions and so are $\tan\beta$
independent. As a result, the $\tan\beta =10$ and $40$ curves lie on top of
one another.  These  three cross sections are
kinematically suppressed because they involve the production of a pair
of
heavy bosons, and  are $\sim 1.5-3$ orders of magnitude below
the dominant $tH^\pm$ cross sections.

In magenta, we show $hH^\pm$ associated production which occurs
dominantly via $s$-channel $W^*$ exchange where the production vertex
includes a factor $g\cos(\alpha +\beta )$ (see Eq.~(8.110) of
Ref. \cite{Baer:2006rs})\footnote{The convention for Higgs mixing angle
  $\alpha$ in Ref. \cite{Baer:2006rs} differs from often-used
  conventions which result in the mixing angle factor
  $\cos(\alpha-\beta)$.} which vanishes in the decoupling limit. As a
result, these cross sections are suppressed from the dominant $tH^\pm$
cross section by $\sim 3-5$ orders of magnitude depending on $m_{H^\pm}$
and $\tan\beta$. They also feature a dip at certain values of
$m_{H^\pm}$ which occurs when the Higgs mixing angle $\alpha$ is such
that $\cos (\alpha +\beta )\to 0$.  An additional cross section $\sigma
(pp\to W^\pm H^\mp+X )$ occurs at the loop level, and so is highly
suppressed and we do not include it here: see Ref. \cite{Dicus:1989vf}.

\begin{figure}[htb!]
\begin{center}
\includegraphics[height=0.4\textheight]{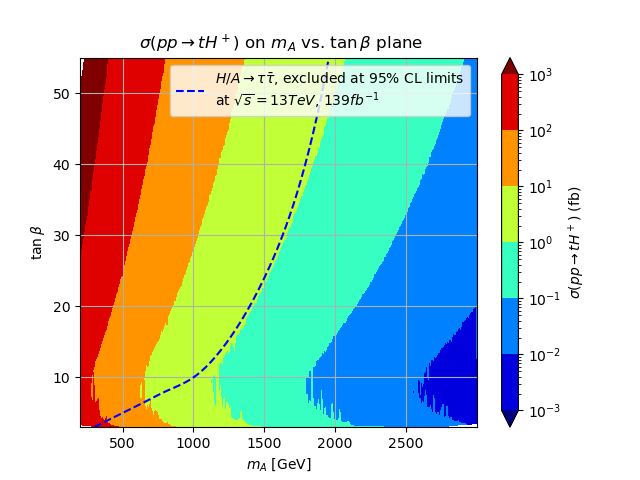}
\caption{The dominant $pp\to tH^\pm +X$ cross section at $\sqrt{s}=14$ TeV
  in the $m_A$ vs. $\tan\beta$ plane.  The region to the left of the
  dashed blue line, is currently excluded at the 95\% confidence level by
  limits from ATLAS searches for $pp\to H,\ A\to\tau^+\tau^-$
  \cite{ATLAS:2020zms}.
\label{fig:sig2}}
\end{center}
\end{figure}

In light of our discussion of the various production cross sections, for
our HL-LHC SUSY charged Higgs reach analysis we will restrict ourselves
to the dominant $tH^\pm$ production process.  In
Fig. \ref{fig:sig2} we show the values of $\sigma (pp\to tH^\pm +X)$ at
LHC14 in the $m_A$ vs. $\tan\beta$ plane for our $m_h^{125}({\rm nat})$
scenario. The largest cross sections $\sim 10^3$ fb are denoted by dark
red whilst the lowest cross sections $\sim 10^{-3}$ fb are denoted dark
blue.  We also show the latest ATLAS 95\% CL exclusion limit from their
search for $H,\ A\to\tau^+\tau^-$ events using LHC13 with 139 fb$^{-1}$
in several non-natural Higgs scenarios which nonetheless maintain
$m_h\sim 125$ GeV\cite{ATLAS:2020zms}.  Thus, to uncover new physics at
HL-LHC in the SUSY Higgs sector, we will mainly focus our attention on
$m_{H^\pm}$ values in the TeV range.

\section{$H^\pm$ branching fractions in natSUSY}
\label{sec:decay}

\begin{figure}[htb!]
\begin{center}
\includegraphics[height=0.28\textheight]{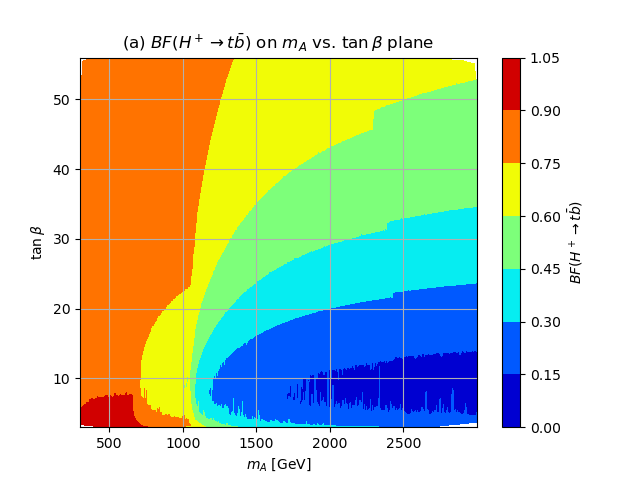 }
\includegraphics[height=0.28\textheight]{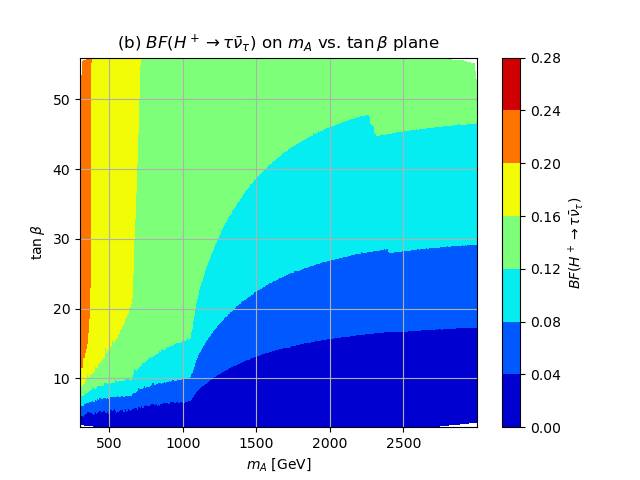}\\
\includegraphics[height=0.28\textheight]{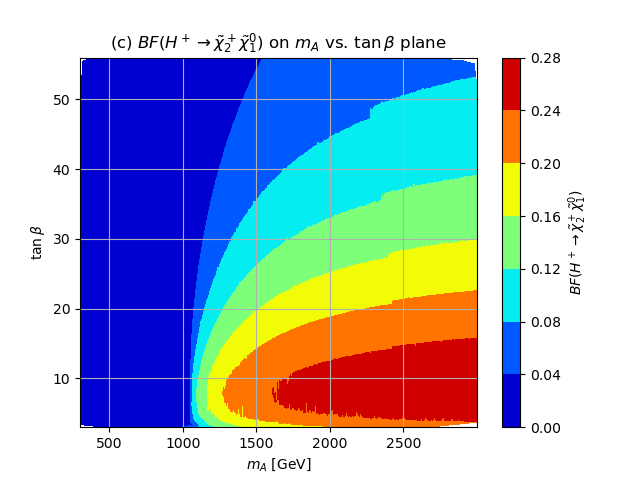}
\includegraphics[height=0.28\textheight]{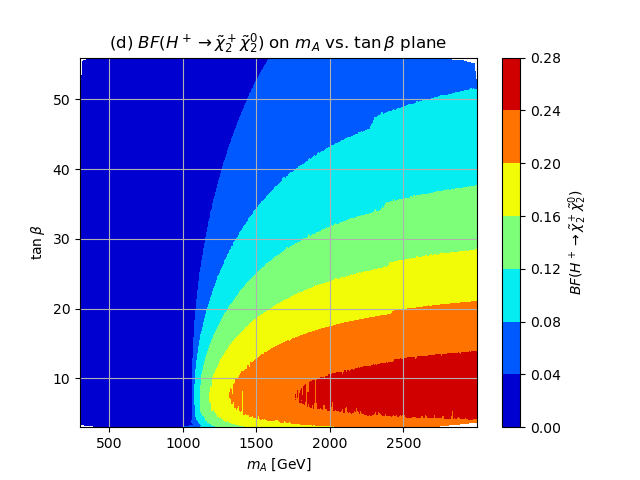}\\
\includegraphics[height=0.28\textheight]{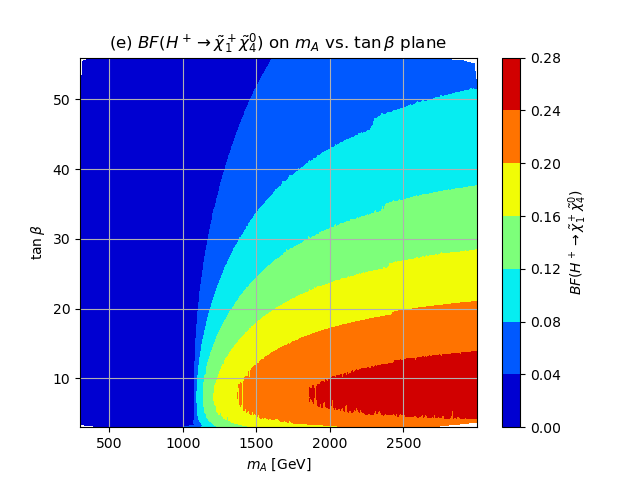}
\includegraphics[height=0.28\textheight]{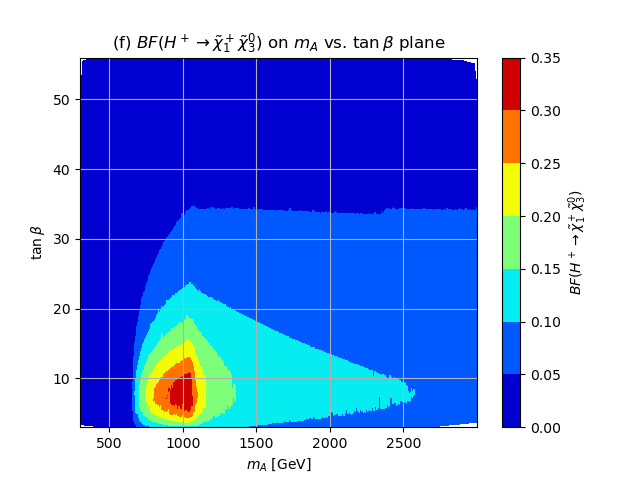}\\
\caption{Branching fractions in the $m_A$ vs. $\tan\beta$ plane for
  $H^+$ to {\it a}) $t\bar{b}$, {\it b}) $\tau^+\nu_{\tau}$, {\it c})
  $\tchi_2^+\tchi_1^0$, {\it d}) $\tchi_2^+\tchi_2^0$, {\t e})
  $\tchi_1^+\tchi_4^0$ and {\it f}) $\tchi_1^+\tchi_3^0$ from Isajet
  7.88\cite{Paige:2003mg} for the model line introduced in the text.
\label{fig:BFHC}}
\end{center}
\end{figure}

The LHC signal from charged Higgs boson production will clearly depend
on how $H^\pm$ decays. For TeV scale values of $m_{H^\pm}$, the dominant SM
decays are via $H^\pm \to tb$ and $H^\pm \to \tau\nu_{\tau}$. Decays to
$h$, $W$ and $Z$ bosons are dynamically suppressed.  Charged Higgs boson
decays via the gaugino plus higgsino modes can also be important if
these are not kinematically suppressed.  With this in mind, in
Fig.~\ref{fig:BFHC}, we show some select $H^{\pm}$ branching fractions
(BFs) in the $m_A$ vs. $\tan\beta$ plane for the model plane
(Eq. \ref{eq:param}).  The branching fractions are color-coded, with the
larger ones denoted by red whilst the smallest ones are denoted by dark
blue.  The branching fractions are extracted from the Isasugra
code\cite{Paige:2003mg}.

In Fig.~\ref{fig:BFHC}{\it a}) we show the BF for $H^+\to t\bar{b}$.
This decay mode to SM particles is indeed dominant for $m_A\alt 1$ TeV
and for larger values of $\tan\beta\agt 20-30$.  In frame {\it b}), we
show the BF($H^+\to\tau^+\nu_{\tau}$).  Like $H^+\to t\bar{b}$, this
mode is enhanced at large $\tan\beta$ and has provided the best avenue
for SUSY charged Higgs discovery/exclusion plots so far.

While SUSY decay modes of $H^\pm$ to higgsino pairs are also open in
these regions, these decay modes are suppressed by mixing angles for
reasons discussed below.
Supersymmetry requires that there is a direct gauge
coupling\cite{Baer:2006rs} \be {\cal L}\ni-\sqrt{2}\sum_{i,A}{\cal
  S}_i^\dagger g t_A\bar{\lambda}_A\psi_i +H.c.  \ee where ${\cal S}_i$
labels various matter and Higgs scalar fields of the MSSM, $\psi_i$ is
the fermionic superpartner of ${\cal S}_i$ and $\lambda_A$ is the
gaugino with gauge index $A$.  Also, $g$ is the corresponding gauge
coupling for the gauge group in question and the $t_A$ are the
corresponding gauge group matrices.  Letting ${\cal S}_i$ be the Higgs
scalar fields, we see there is an unsuppressed coupling of the Higgs
scalars to a gaugino and a  higgsino as mentioned earlier.  This coupling
can lead to dominant SUSY Higgs boson decays to SUSY particles when the
gaugino-plus-higgsino decay channel is kinematically unsuppressed. But
it also shows why the heavy Higgs decay to higgsino pairs is suppressed
by mixing angles for $|\mu| \ll |M_{1,2}|$, once we recognize that a
Higgs boson-higgsino-higgsino coupling is forbidden by gauge invariance.

In frame {\it c}), we show BF($H^+\to \tchi_1^0\tchi_2^+$), where
$\tchi_1^0$ is dominantly higgsino-like and $\tchi_2^+$ is dominantly
wino-like for natural SUSY models like the $m_h^{125}({\rm nat})$
scenario. Here, we see that for larger values of $m_A\simeq m_{H^+}\agt
1.2$ TeV, then this mode turns on, and at least for moderate
$\tan\beta\sim 10-20$ (which is favored by
naturalness\cite{Bae:2014fsa}), rapidly comes to dominate the $H^+$
decay modes along with the neutral wino+higgsino channels $H\to
\tchi_2^+\tchi_2^0$ (frame {\it d})) and $H^+\to \tchi_1^+\tchi_4^0$
(frame {\it e})). In  our analysis we have assumed that the
  gaugino mass parameters unify at the high scale, so that at the weak
  scale $M_1 \simeq {1\over 2}M_2$: as a result,  $\chi_3^0$ is dominantly
  bino-like, $\chi_4^0$ and $\chi_2^\pm$ are dominantly wino-like,
while $\tchi_{1,2}^0$ and $\chi_1^\pm$ are mainly
higgsino-like. The sum of these three wino+higgsino decay channels thus
dominate the $H^+$ decay branching fractions for $m_{H,A}\agt 1.2$ TeV
and low-to-moderate values of $\tan\beta$. For high values of
$\tan\beta$, the $b$ and $\tau$ Yukawa couplings become large, and SM
decays to fermions once again dominate SUSY decays. Decays of $H^+$ to
gauge boson pairs and to $h$ are unimportant in the decoupling limit as
mentioned above.  For
completeness, we also show in frame {\it f}) the decay mode
$H\to\tchi_1^+\tchi_3^0$ which is to higgsino+bino.  This mode is large
only in a small region of $m_H\sim 1$ TeV and modest $\tan\beta$ where
the mode $H\to bino+higgsino$ decay has turned on, but where $H\to
wino+higgsino$ has yet to become kinematically open. Decays to winos
dominate decays to binos because the $SU(2)$ gauge coupling $g$ is
larger than the hypercharge gauge coupling $g^\prime$.

\section{Search for $H^{\pm} \to \tau \nu_{\tau}$}
\label{sec:taunu}

Next, we turn to the examination of the prospects for discovering the
charged Higgs boson produced
at the HL-LHC via $pp\to H^\pm t+X$,  followed by
$H^\pm\to \tau\nu_\tau$ decay.
For signal and $2\to 2$ background processes listed below,
we use the Pythia event generator\cite{Sjostrand:2007gs}.
For $2\to 3$ BG processes such as
$Wb\bar{b}$, $Zb\bar{b}$ and $ht\bar{t}$ production, we adopt
Madgraph\cite{Alwall:2011uj} for the subprocess calculation but then
interface with Pythia for parton showers, hadronization and underlying event.
Our final state particles are then fed into the
Delphes\cite{deFavereau:2013fsa} detector simulation program
which includes a jet-finding algorithm and routines for identifying
both $b$-jets and hadronic tau jets (labelled as $\tau_h$).

In Delphes, a jet object is reconstructed using an anti-$k$ algorithm with
$p_T(min) > 25$ GeV and $\Delta R < 0.4$.
For a baseline jet we require: 
\bi
\item $|\eta (j)| < 4.7$.
  \ei
  For a baseline $b$-jet, besides the requirement for a baseline jet,
  we further require
  \bi
\item the jet to be tagged as a $b$-jet by Delphes.
  \ei
  
  For a signal $\tau_h$-jet, besides the requirement of baseline jet,
  we further require
  \bi
  \item the jet must be a tagged hadronic tau-jet $\tau_h$ by Delphes with
  \item $|\eta (\tau_h )| < 2.5$.
    \ei
    
For the baseline lepton isolation requirement, we require
\bi
\item $p_T(min)(e/\mu ) > 5$ GeV.
  \ei
  For a signal lepton, besides the requirement for baseline lepton isolation,
  we further require
  \bi
\item $|\eta_{e,(\mu )}| < 2.47\ (2.5)$ and
  \item $p_T(e(\mu)) > 20\ (25)$ GeV.
  \ei

\subsection{$H^\pm\to\tau\nu_\tau\to \tau_h+\eslt$ channel}
\label{ssec:tau}

In this channel, we search for $H^\pm\to\tau_h+\eslt$ along with the
presence of a spectator $t$-jet which is signalled by the presence of
a tagged $b$-jet. We include
SM BGs from $t\bar{t}$, single top, $Wb\bar{b}$, $Zb\bar{b}$,
$WW$, $WZ$, $ZZ$, $Zh$, $Wh$ and $ht\bar{t}$ production.
We first require:
\bi
\item exactly one signal $\tau_h$-jet with no baseline leptons; the no
  baseline lepton requirement
  targets events where the spectator top decays hadronically, though of
  course events with a semileptonic decaying top could contribute if
  the lepton evades detection;
\item $n(b-{\rm jet})\ge 1$, where $b$ here (and in the rest of 
  Sec.~\ref{sec:taunu}) refers to baseline $b$-jets,  and
\item a neutrino reconstruction method\footnote{In $t\bar{t}$ events
  where one of the tops decays hadronically, and the other leptonically,
  so that the $\eslt$ comes only from the (massless) neutrino;
  i.e. $\eslt_{x,y}= p(\nu)_{x,y}$, one can construct $p(\nu)_z$ assuming
  that the $W$ boson from top decay is on-shell. This vetoes about half
  the potentially enormous $t\bar{t}$ background with a loss of less
  than 10\% of the signal.} is employed here.  If the invariant mass of
  the reconstructed neutrino, the signal $\tau_h$-jet and any of the tagged baseline
  $b$-jets in the event reconstructs to $m=m_t\pm 50$ GeV, then the
  event is vetoed.  \ei This latter requirement is
  imposed to veto a portion of the very large $pp\to t\bar{t}$ background.

  With these remaining events, an examination of various distributions
  of signal and background (that we do not show for brevity) leads us to
  impose the following analysis cuts:
  \bi
\item $\eslt > 350$ GeV,
  \item $\Delta\phi (\tau_h ,\vec{\eslt}) > 30^\circ$,
  \item $\Delta\phi (b_1,\vec{\eslt}) > 50^\circ$, where $b_1$ is the leading baseline
    $b$-jet, and
  \item $min(\Delta R(b,\tau_h )) > 0.9$, where the $b$ loops over all tagged
    baseline  $b$-jets
    in the event.
    \ei

\begin{figure}[htb!]
\begin{center}
\includegraphics[height=0.5\textheight]{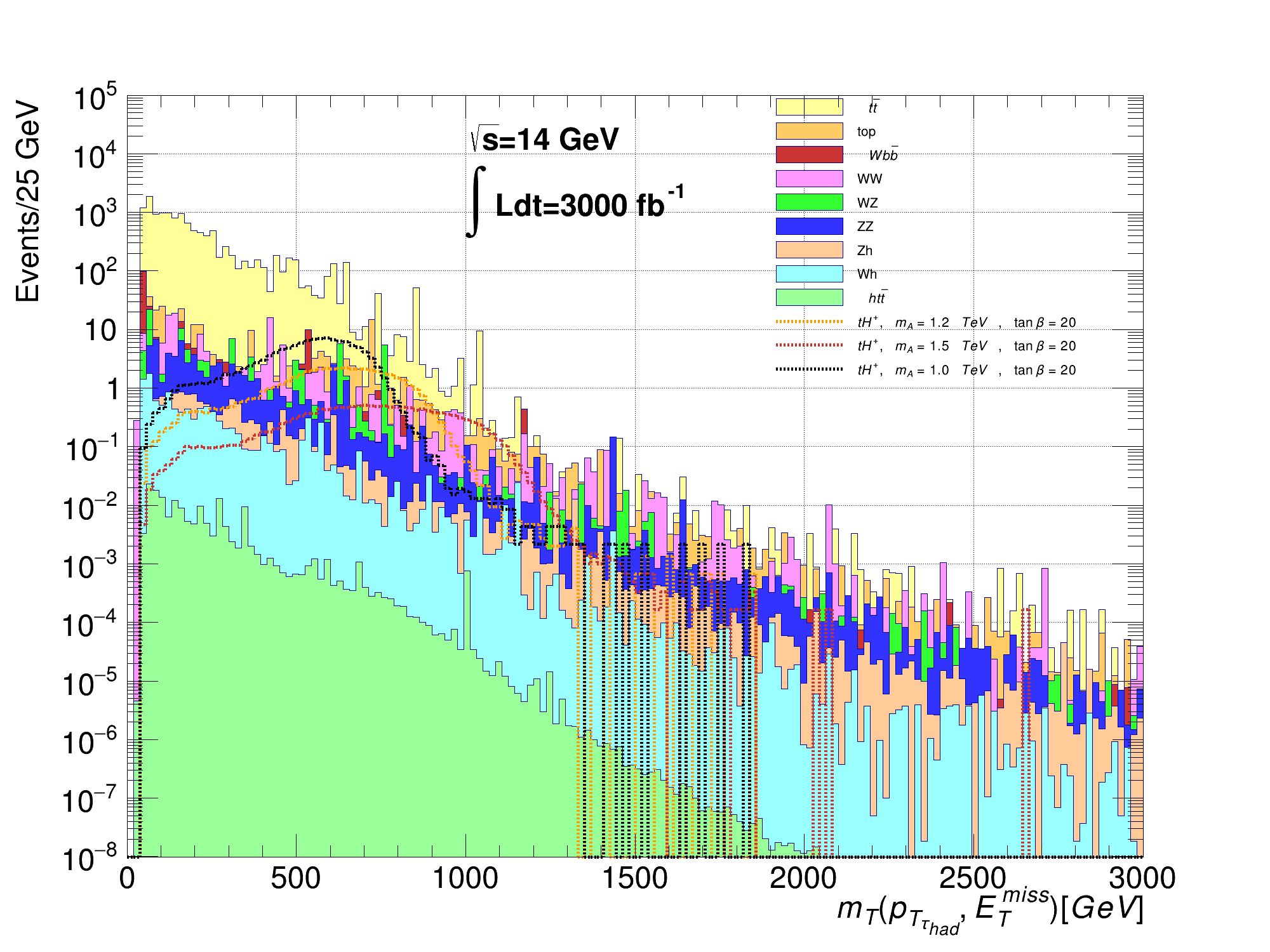}
\caption{Distribution in $m_T(\tau_h,\eslt )$ from
  $pp\to H^\pm(\to \tau\nu)+ t+X$  followed by the decay
  $\tau\to hadrons +\eslt$.
  We also show dominant SM backgrounds.
\label{fig:mTtau}}
\end{center}
\end{figure}

    After these cuts, the resulting transverse mass distribution
    $m_T(\tau_h,\eslt )$ is shown in Fig. \ref{fig:mTtau}. As expected,
    the signal histograms peak around the value of $m_{H^\pm}$, while
    the backgrounds yield falling distributions. 
    Our goal in each signal channel is to look for an excess above the SM
    backgrounds in the largest transverse mass bins which are most sensitive
    to TeV-scale charged Higgs decay.
    From the distribution, the solid colored histograms represent the
    various BGs, of which the dominant is light yellow: $t\bar{t}$.
    The signal distributions, labelled as dotted curves for several
    benchmark scenarios as listed, can emerge from BG at large values of
    $m_T$, provided the signal is large enough.

\subsection{$H^\pm\to\tau\nu_\tau\to \ell+\eslt$  channel}
\label{ssec:l}

In this subsection, we examine the $pp\to tH^\pm+X$ production
reaction where $H^\pm\to\tau\nu_\tau$ followed by
$\tau\to\nu_\tau+(\ell\nu_\ell )$ channel, where $\ell =e$ or $\mu$.
For this signal channel, we require:
\bi
\item exactly one signal lepton and no other baseline leptons,
\item no jets have been $\tau$-tagged ($\tau_h$-veto); here, we are again
  targeting events where the top decays hadronically, and the tau
  decays leptonically.
\item $n(b-{\rm jet})\ge 1$ and
\item the neutrino reconstruction method described above is employed here.
  If the invariant mass of the reconstructed neutrino,
  the signal lepton plus any of the baseline $b$-jets in the event is within
  $m=m_t\pm 50$ GeV, then event is vetoed.
\ei
Standard Model backgrounds from $t\bar{t}$, single top, $Wb\bar{b}$,
$WW$, $WZ$, $Wh$, $ht\bar{t}$ and $Zh$ can also lead to the same event
topology as the signal. 

\begin{figure}[htb!]
\begin{center}
\includegraphics[height=0.5\textheight]{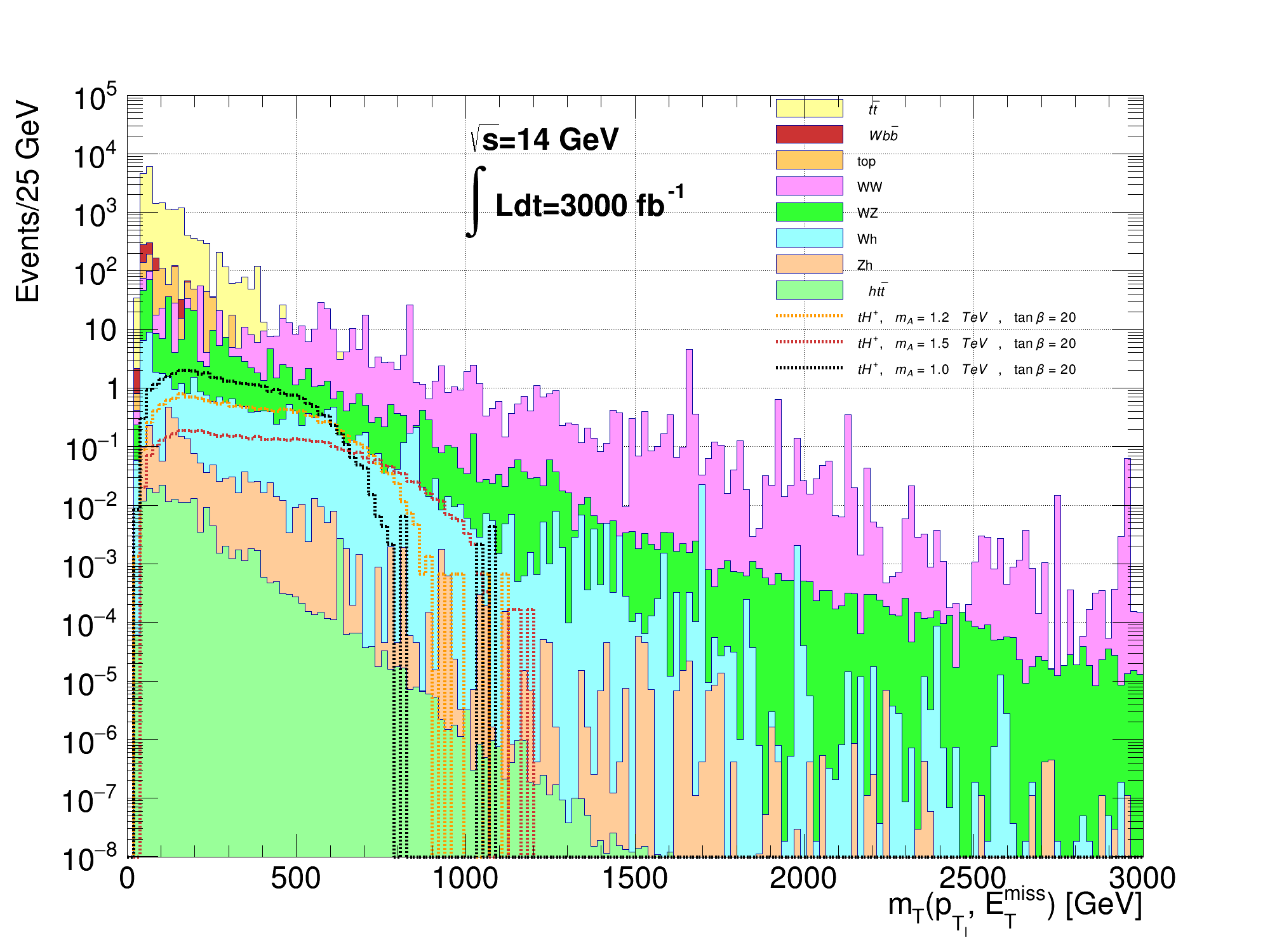}
\caption{Distribution in $m_T(\ell,\eslt )$ from $pp\to H^\pm(\tau\nu)+
  t + X$
  followed by $\tau\to\ell+\eslt$ decay. We also show
  dominant SM backgrounds.
\label{fig:mT1l}}
\end{center}
\end{figure}

Examination of various distributions leads us to impose the following
analysis cuts:
\bi
\item $\eslt > 350$ GeV,

\item $\eslt_{,rel}:=$
  $\eslt\cdot\sin{(min(\Delta\phi,\frac{\pi}{2}))}>150$ GeV, where
  $\Delta\phi$ is the azimuthal angle between the $\Vec{\eslt}$ and the
  closest lepton or jet with $p_T > 25$ GeV.
\item $\Delta\phi (\ell,\vec{\eslt}) > 30^\circ$,
\item $\Delta\phi (b_1,\vec{\eslt}) > 70^\circ$, where $b_1$ is the leading baseline
  $b$-jet and
\item $min(\Delta R(b,\ell)) > 1.2$, where the $b$ loops over all tagged baseline
  $b$-jets in the event.
\ei
The resultant $m_T(\ell,\eslt )$ distribution is displayed in
Fig. \ref{fig:mT1l}. The dominant BG at low $m_T$ comes from $t\bar{t}$
production (light-yellow histogram) while the dominant BG at high $m_T$
comes from $WW$ production. We also show several signal benchmark
distributions (dotted curves) which may cause an excess of events
over background expectations at high $m_T(\ell,\eslt )$.
In this case, the signal distributions shown will only cause a slight excess
above background at high $m_T$. But combined with the other channels,
this signal channel can slightly increase the overall significance.

\subsection{$H^\pm\to\tau\nu_\tau$ with $t\to b\ell\nu_\ell$ channel}
\label{ssec:ltau}

In this channel, we attempt to extract the signal from $pp\to tH^\pm$
production followed by $H^\pm \to\tau_h+\eslt$ but where the spectator
$t$-quark decays semi-leptonically: $t\to b\ell\nu_\ell$.
SM backgrounds from  $t\bar{t}$, single top, $WW$, $WZ$, $Wh$, $Zh$
and $ht\bar{t}$ production are included in our analysis.

We require:
\bi
\item exactly one signal lepton and no other baseline leptons,
\item exactly one signal $\tau_h$-jet,
\item the charges of the signal lepton and the $\tau_h$-jet
  must be OS (opposite-sign) and
\item $n(b-{\rm jet})\ge 1$.
\ei
  
  Examination of the resultant signal and background distributions
  leads us to the following additional analysis cuts:
  \bi
\item $\eslt > 350$ GeV,
\item $\eslt_{rel} > 150$ GeV,
\item $\Delta\phi(\ell,\vec{\eslt}) > 20^\circ$,
\item $\Delta\phi (\tau_h ,\vec{\eslt}) > 80^\circ$,
\item $\Delta\phi (b_1,\vec{\eslt}) > 50^\circ$, where $b_1$ is the leading baseline
  $b$-jet and
\item $min(\Delta R(b,\tau_h )) > 1.2$, where the $b$ loops over all baseline $b$-jets
  in the event.
  \ei

\begin{figure}[htb!]
\begin{center}
\includegraphics[height=0.5\textheight]{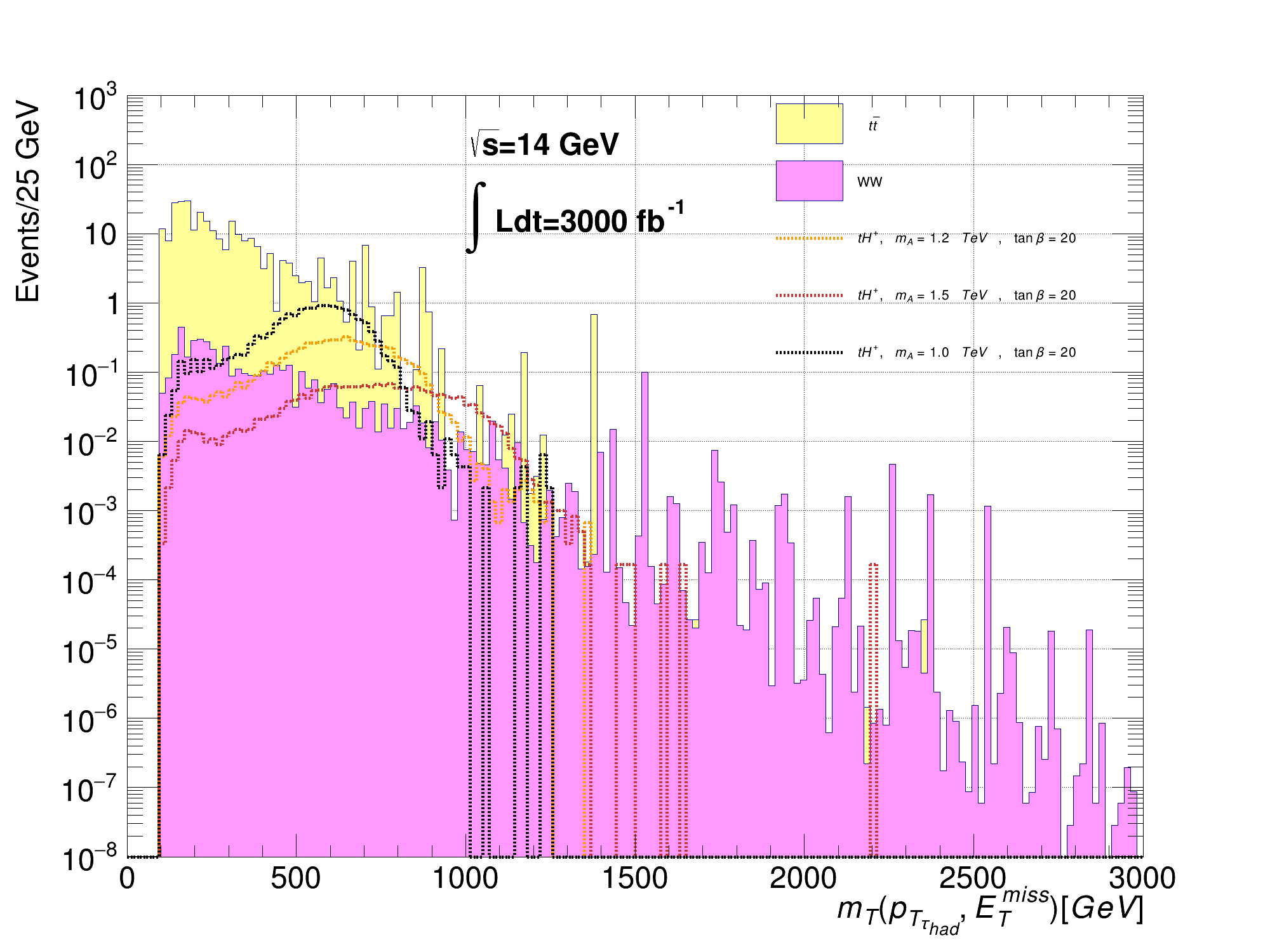}
\caption{Distribution in $m_T(\tau_h,\eslt )$ from $pp\to H^\pm t + X$
  followed by $H^\pm\to \tau_h+\eslt$ decay and also $t\to b\ell\nu_\ell$
  decay. We also show dominant SM backgrounds.
\label{fig:mTltau}}
\end{center}
\end{figure}

  The resultant $m_T(\tau_h,\eslt )$ distribution is shown in
  Fig. \ref{fig:mTltau}. The dominant BG at low $m_T$ again comes from
  $t\bar{t}$ production. At high $m_T$, then the various signal
  histograms can cause a noticeable increase in the expected $m_T$
  distribution beyond SM expectations, albeit with a rather low event rate.

\subsection{LHC reach in $H^\pm\to\tau\nu_\tau$ channel}
\label{ssec:tau_reach}

Using the analysis cuts for the various signal channels delineated
above, we can now create reach plots to show the LHC14 discovery sensitivity
or exclusion limits for $pp\to H^\pm t$ production in the
$m_A$ vs. $\tan\beta$ plane.  We use the $5\sigma$ level to claim
discovery of a 
charged Higgs boson and assume the true distribution one observes
experimentally corresponds to signal-plus-background.  We then test this
against the background-only distribution in order to see if the
background-only hypothesis can be rejected at the $5\sigma$ level.
Specifically, we use the binned transverse mass distributions (bin width
of 25 GeV) from each signal channel as displayed above to obtain the
discovery/exclusion limits.

In the case of the exclusion plane, the upper limits for exclusion
of a signal are set at 95\% CL; one assumes the true distribution
one observes in experiment corresponds to background-only.
The limits are then computed using a modified frequentist $CL_s$
method\cite{Read_2002} where the profile likelihood ratio is
the test statistic. 
In both the exclusion and discovery planes, the asymptotic approximation
for obtaining the median significance is employed\cite{Cowan_2011}.

\begin{figure}[htb!]
\begin{center}
  \includegraphics[height=0.4\textheight]{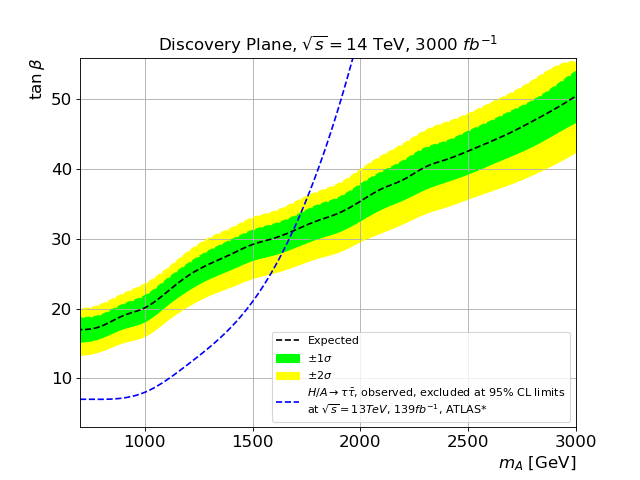}\\
  \includegraphics[height=0.4\textheight]{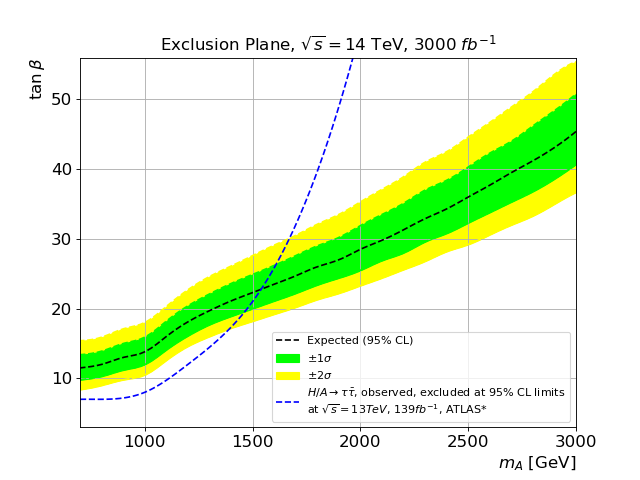}\\
  \caption{In {\it a}), we show the $5\sigma$ discovery region of the
    $m_A$ vs. $\tan\beta$ plane for $pp\to H^\pm t+X$
    followed by $\tau\to\ell+\eslt$ decay for HL-LHC with 3000 fb$^{-1}$.
    In {\it b}), we plot the corresponding 95\% CL exclusion region. The
    region above the dashed blue curve is excluded by ATLAS searches for
    $H/A \to \tau\tau$ events.    
  \label{fig:disc3000taunu}}
\end{center}
\end{figure}

In Fig. \ref{fig:disc3000taunu}, we plot our result for the
discovery/exclusion regions via the $H^\pm \to \tau\nu$ channel for the
HL-LHC with $\sqrt{s}=14$ TeV and 3000 fb$^{-1}$ of integrated
luminosity in the $m_A$ vs. $\tan\beta$ plane using our $m_h^{125}({\rm
  nat})$ benchmark scenario (which is quite typical for natural SUSY
models\cite{Baer:2020kwz}).  In frame {\it a}), we plot the $5\sigma$
discovery reach using the combined three channels discussed previously. Above
the dashed black line, experiments at the HL-LHC should be able to
discover $H^\pm$ at the LHC operating at $\sqrt{s}=14$~TeV, assuming an
integrated luminosity of 3000~fb$^{-1}$.  The green and yellow bands
display the $\pm 1\sigma$ and $\pm 2\sigma$ uncertainties in our mapping
of the discovery region. The region above the dashed blue line is
excluded by ATLAS searches for $H/A \to \tau\tau$ events, albeit in a
scenario with decoupled superpartners \cite{ATLAS:2020zms}.  From the
plot, we see that a discovery region does indeed emerge, starting around
$m_A\sim 500$ GeV and $\tan\beta \sim 18$ and extends out to $m_A\sim 3$
TeV for $\tan\beta\sim 50$ where both the $\sigma(pp\to tH^\pm +X)$ and the
branching fraction for $H^+\to \tau\nu_\tau$ decays are both enhanced.
The discovery region pinches off below $\tan\beta\sim 15$ where the
$H^\pm\to\tau\nu_\tau$ branching ratio becomes too small.

In frame {\it b}), we plot the 95\%
CL exclusion limit for HL-LHC for our combined three signal channels.
The exclusion limit now extends out to beyond $m_A\sim 3$ TeV for large
$\tan\beta\sim 50$.
We also see that the exclusion contour extends to about
$\tan\beta\sim 10$ for relatively light $m_{H^+}$; however, this
part of the plane is already excluded by ATLAS searches.

\section{Search for $H^\pm\to tb$}
\label{sec:tb}

In this section, we examine the capability of HL-LHC to discover charged
Higgs bosons in the $H^\pm\to tb$ decay channel. Recent limits have been placed
within this search channel by the ATLAS collaboration using 139 fb$^{-1}$
of data\cite{ATLAS:2021upq}.
Our analysis proceeds similarly to Sec. \ref{sec:taunu} except now we place an
emphasis on the presence of high-$p_T$  top-jets in the final state.

In the following, we will use lower case letter such as $j$ ($b$)
to denote the small radius jets (tagged $b$-jets) while upper case letter
such as $J$ ($T$) denote large radius jets (top-jets). 

The parameters for baseline reconstructed particles are, for
charged leptons $l$:
\bi
\item $|\eta (e, \mu )| < 4$,
\item $p_T(e) > 20$ GeV,\ \ \ $p_T(\mu) > 25$ GeV.
  \ei
For small radius jets $j$:
\bi
\item jet reconstruction with the anti-$k_T$ algorithm,
\item cone size $\Delta R < 0.4$,
\item $p_T(j) > 25$ GeV,
\item $|\eta (j)| < 4.7$.
  \ei
For a tagged baseline $b$-jet:
\bi
\item satisfy the above small radius jet requirements,
\item tagged $b$-jet by Delphes.
  \ei
  For a large radius jet $J$:
  \bi
\item jet reconstruction using the Cambridge/Aachen algorithm (CA)\cite{CMS:2009lxa,Anders:2013oga},
\item cone size $\Delta R < 1.5$,
\item $p_T(J) > 300$ GeV and
\item $|\eta (J)| < 2$.
  \ei
  For a tagged top-jet $T$:
  \bi
\item satisfy the above large radius jet ($J$) requirements,
\item $T$ is tagged by the HEPTopTagger2 algorithm\cite{Plehn:2011sj,Anders:2013oga}.
  \ei
  Also, for candidate events with an isolated lepton or a signal $b$-jet,
  we further require
  \bi
  \item $|\eta (e)| < 2.47$, $|\eta (\mu)| < 2.5$ and $|\eta (b)| < 2.4$.
\ei

The analysis is then separated into four orthogonal channels depending on
whether or not the final state does or does not contain a tagged $T$-jet, and
whether or not it contains an isolated lepton.

In all cases, the small radius $b$-jet (denoted as $b_1$ below)
arising directly from the $H^\pm$ decay is determined by the following
procedure. 
\bi
\item we require $p_T(b_1) > 350$ GeV,
\item $|\eta (b_1)| < 1.5$,
\item $R(b_1, J_1) > 1.5$ (so this means $b_1$ must be outside the cone of the
  fat jet top candidate).
\item $m(j, j^\prime , b_1)$ {\it cannot} be in the top mass range
  $[125, 225]$ GeV, where $j$, $j^\prime$ are any small radius jet pairs
  in the event.
\item If multiple candidates satisfying these conditions are found,
  the one with the hardest $p_T$ is taken as $b_1$.
\item Events are vetoed if no $b$-jets satisfy these conditions.
\ei
Then, $m(T_1, b_1)$ is used to reconstruct the mass of the $H^\pm$
in all cases. Note that in the channels where the HEPTopTagger2 has
positively tagged a top-jet, {\it i.e.} the $1t, 1t1l$ channels,
it is the four vector reconstructed by the algorithm that
is taken as $T_1$. But in the channels where HEPTopTagger2 fails to
identify a top-jet,
then it is the fat jet itself that is taken as the $T_1$.
The $m(T_1, b_1)$ distributions are then shown as the final results for
each signal channel.

The background samples being considered for the hadronic channels $[1t,
  1t(no \ tag)]$ are $t\bar{t}$ (with $\ge 3$ truth $b$'s removed to
avoid double counting with $t\bar{t}b\bar{b}$ events that are separately
simulated), single top, $t\bar{t}b\bar{b}$ and $tttt$.  The backgrounds
for the semileptonic channels $[1t1l, 1t(no \ tag)1l]$ are $t\bar{t}$
(again with $\ge 3$ truth $b$s removed), $t\bar{t}b\bar{b}$ and
$t\bar{t}t\bar{t}$. We have not simulated $t\bar{t}V$ events as we
require at least three $b$-jets which has been shown to be very small
\cite{ATLAS:2018ntn}.

\subsection{Single tagged top channel without signal leptons}

In this channel, we search for $pp\to tH^\pm+X$ with $H^\pm\to tb$
decay.  The primary $t$-quark in the $t H^\pm$ final state tends to be
non-central, and so rarely produces a tagged top-jet, but does more
often produce a tagged $b$-jet. So for this channel, we focus on
reconstructing the decay-produced top jet from $H^\pm \to tb$, where
TeV-scale charged Higgs decay gives rise to a well-collimated
$T$-jet. Thus, \bi
\item The HEPTopTagger2 algorithm  that we use has tagged exactly one
  top from the large radius boosted ($R < 1.5$, $p_T(J) > 300$ GeV) jet
  $n_T = 1$. (This fat jet is denoted as $J_1$ below).
  \ei
  The top-jet four vector reconstructed by the HEPTopTagger2 is used as $T_1$.
  The four vector for the subjet $b$ reconstructed by the tagger is
  denoted as $b_2$. We further require the following.
  \bi
\item At least 3 $b$-jets: $n_b\ge 3$, of which at least two of them
  must satisfy the signal $b$-jet requirements listed above.
\item At least 6 jets: $n_j\ge 6$.
\item No isolated leptons: $n_l = 0$.
\ei

Based on examination of various signal/background distributions, we also require
\bi
\item $H_T > 1200$ GeV,
\item $m(b, b^\prime) > 215$ GeV, where the $b$ and $b^\prime$ are the
  $b$-jet pair with the max $p_T$ in the events,
\item $max(R(b^\prime, H^\pm)) > 1.5$, where the $b^\prime$ are any $b$-jets
  in events that are not $b_1$ and $b_2$. The $H^\pm$ is reconstructed from
  $T_1$, $b_1$,
\item $min(R(b^\prime , b_1)) < 2.8$, where $b^\prime$ are any $b$-jets
  in the events.
\ei

\begin{figure}[htb!]
\begin{center}
\includegraphics[height=0.5\textheight]{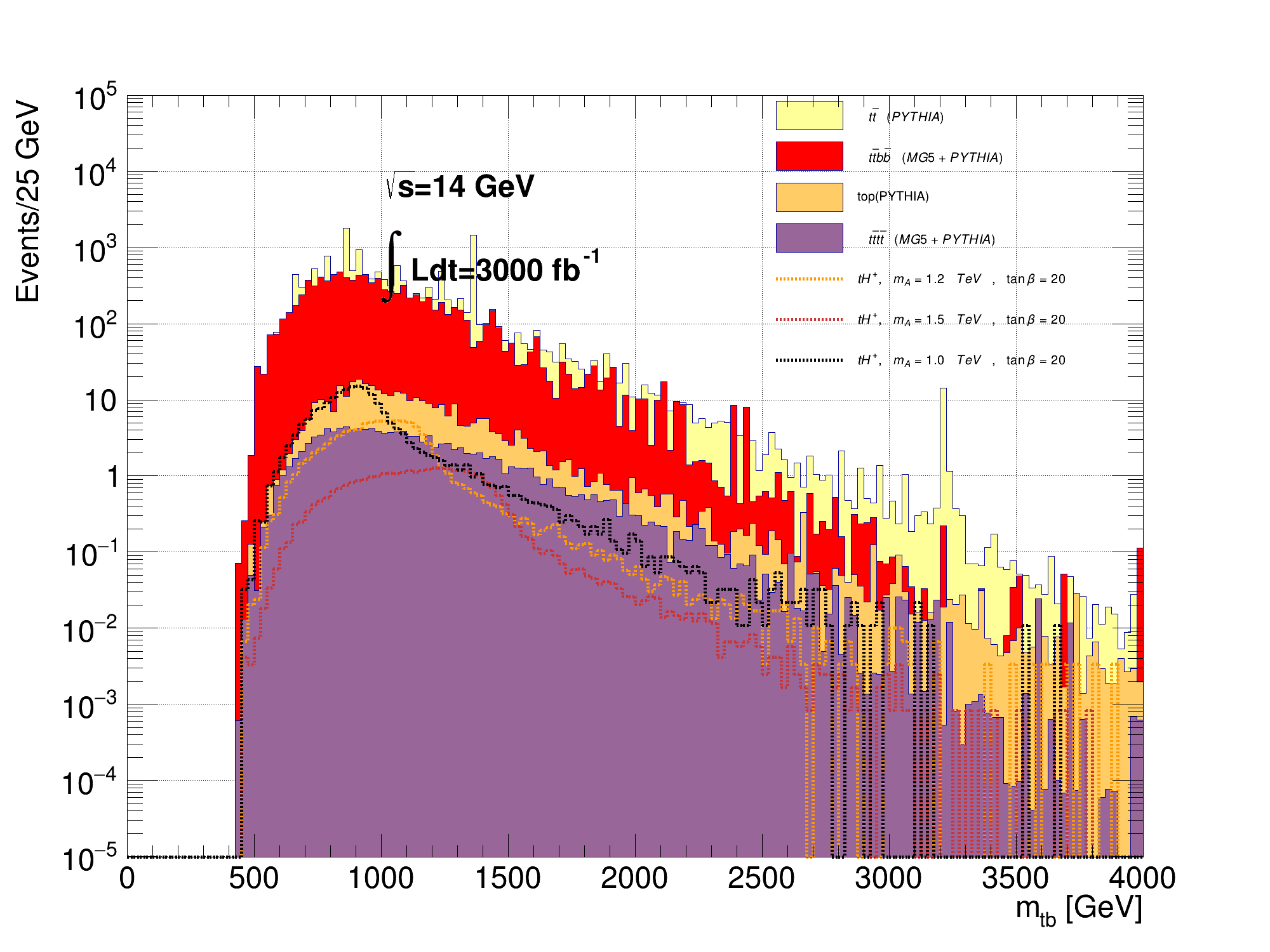}
\caption{Distribution in $m(tb)$ from $pp\to H^\pm t+X$
  followed by $H^\pm\to tb$ decay.
  These events include a  top-jet tagged by HEPTopTagger2.
  We also show dominant SM backgrounds. 
\label{fig:mtb_1t}}
\end{center}
\end{figure}

At this point, we are able to construct the distribution $m(tb)\equiv
m(T_1,b_1)$ shown in Fig. \ref{fig:mtb_1t}. The dominant background
distributions are shown as solid colored histograms whilst several
signal benchmark models are shown as dashed histograms. From the plot,
we see that the signal distributions roughly reconstruct $m(H^\pm )$
while background is dominated by $t\bar{t}b\bar{b}$ production at low
mass, and by $t\bar{t}$ at high invariant mass. The goal then is to
search for resonant signal bumps against the continuum of expected
backgrounds. If this bump is buried under the SM background, this channel
will make a negligible contribution to the significance
when combined with other channels.

\subsection{Single top (no tag) channel without signal leptons}

In this channel, the HEPTopTagger2 fails to tag any tops.
However,
\bi
\item there is at least one large radius boosted ($R < 1.5$, $p_T > 300$
  GeV) jet $n_J >= 1$ with trimmed mass\cite{Krohn_jet_trim} $115\ {\rm GeV} < m_J < 190$ GeV, and at
  least one small radius ($R<0.4$) $b$-jet within the cone of the large
  radius jet $J$.  Then, the fat Jet with the hardest $p_T$ is taken as
  the top candidate arising directly from the charged Higgs decay
  (denoted as $J_1/T_1$).  The hardest $b$ within $J_1/T_1$ cone is
  taken as $b_2$.
\item At least 3 $b$ jets: $n_b\ge  3$, of which at least two of them
  must satisfy the signal $b$-jet requirements listed above.
\item At least 6 jets: $n_j\ge 6$ and
\item No isolated leptons: $n_l = 0$.
\ei
  
We then require:
\bi
\item $H_T > 1200$ GeV,
\item $m(b, b^\prime ) > 215$ GeV, where the $b$ $b^\prime$ are the $b$-jet
  pairs with the max $p_T$ in the events,
\item $max(R(b^\prime , H^\pm )) > 2.4$, where the  $b^\prime$ are any $b$-jets
  in events that are not $b_1$ and $b_2$.
  The $H^\pm$ is reconstructed from $T_1$, $b_1$,
\item $ 0.8 < min(R(b^\prime , b_1)) < 2.8$, where the $b^\prime$ are any
  $b$- jets in the events.
\ei

\begin{figure}[htb!]
\begin{center}
\includegraphics[height=0.5\textheight]{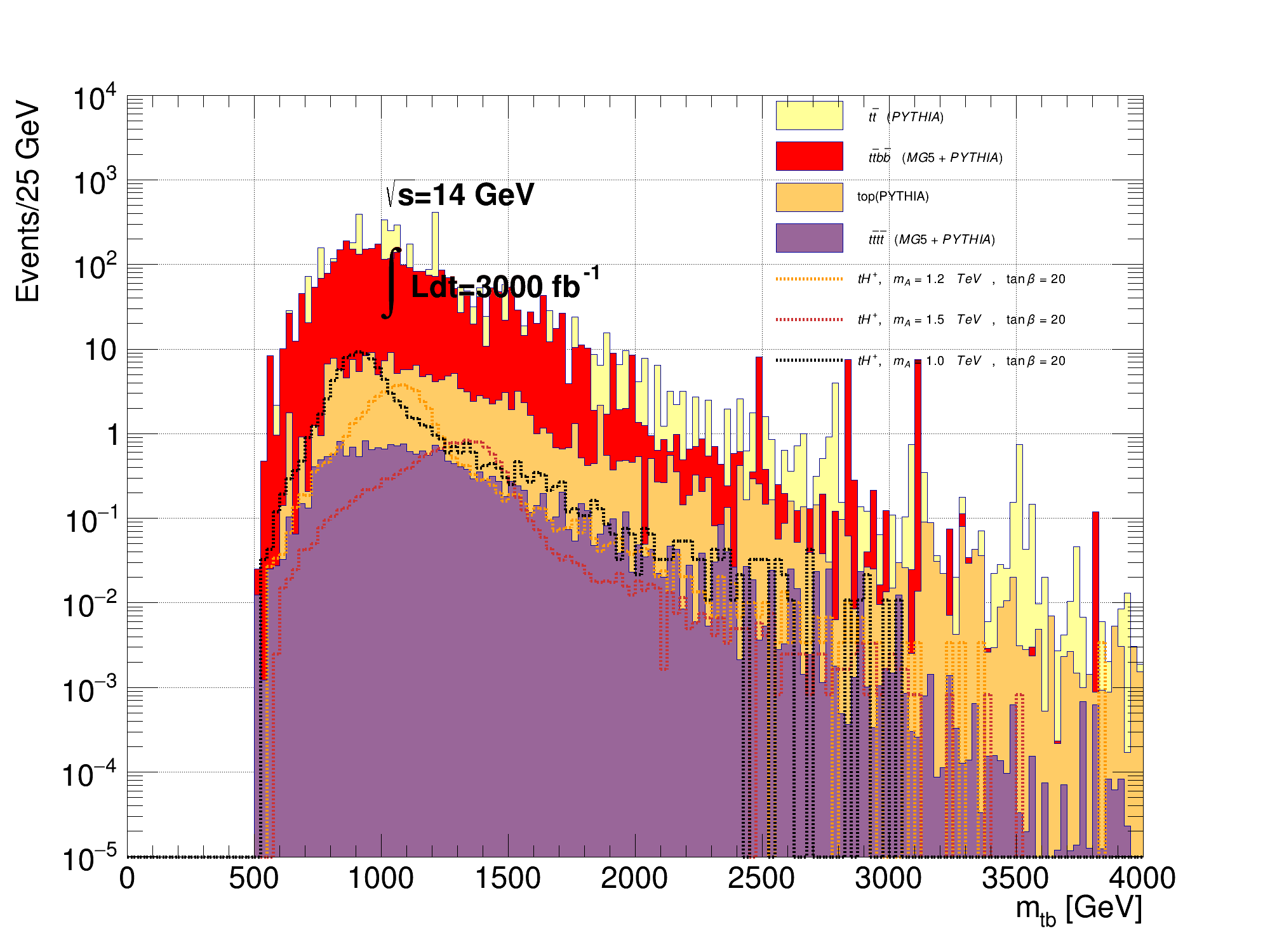}
\caption{Distribution in $m(tb)$ from $pp\to H^\pm t+X$
  followed by $H^\pm\to tb$ decay, but no fat jet tagged as top
  by HEPTopTagger2. We also show dominant SM backgrounds.
\label{fig:mtb_1t_notag}}
\end{center}
\end{figure}

The distribution $m(tb)$, reconstructed from $T_1$ and $b_1$, is shown
in Fig. \ref{fig:mtb_1t_notag} where again the $m(tb)$ roughly
reconstructs $m(H^\pm )$ and where again the $t\bar{t}$ and
$t\bar{t}b\bar{b}$ are the dominant backgrounds.

\subsection{Single top (tagged) plus lepton channel}

In this signal channel, we again require tagged top-jet, but now also
require the presence of an isolated lepton arising from semileptonic
decay of one of the tops.

We require
\bi
\item The HEPTopTagger2 has tagged exactly one top from the large radius
  boosted ($R < 1.5$, $p_T > 300$ GeV) jet $n_T= 1$.
 
 As before the top four vector reconstructed by the HEPTopTagger2 is
 denoted as $T_1$.  The four vector for the subjet $b$ reconstructed by
 the tagger is denoted as $b_2$.

\item At least 3 $b$-jets: $n_b\ge 3$, of which at least two of them
  must satisfy the signal $b$-jet requirements listed above.
\item At least 4 jets: $n_j\ge 4$.
\item Exactly one signal isolated leptons: $n_l = 1$.
  \ei

  We further require
  \bi
  \item $H_T > 1200$ GeV,
  \item $m(b, b^\prime ) > 215$ GeV, where the $b$,  $b^\prime$ are the $b$
    pairs with the max $p_T$ in the events,
  \item $max(R(b^\prime ,H^\pm )) > 1.5$, where $b^\prime$ are any $b$-jets
    in the events that are not $b_1$ and $b_2$.
    The $H^\pm$ is reconstructed from $T_1$ and $b_1$.
  \item $min(R(b^\prime , b_1)) > 1$, where the $b^\prime$ are any $b$-jets
    in the events, and
  \item $R(b_1,l) > 0.9$.
    \ei

\begin{figure}[htb!]
\begin{center}
\includegraphics[height=0.5\textheight]{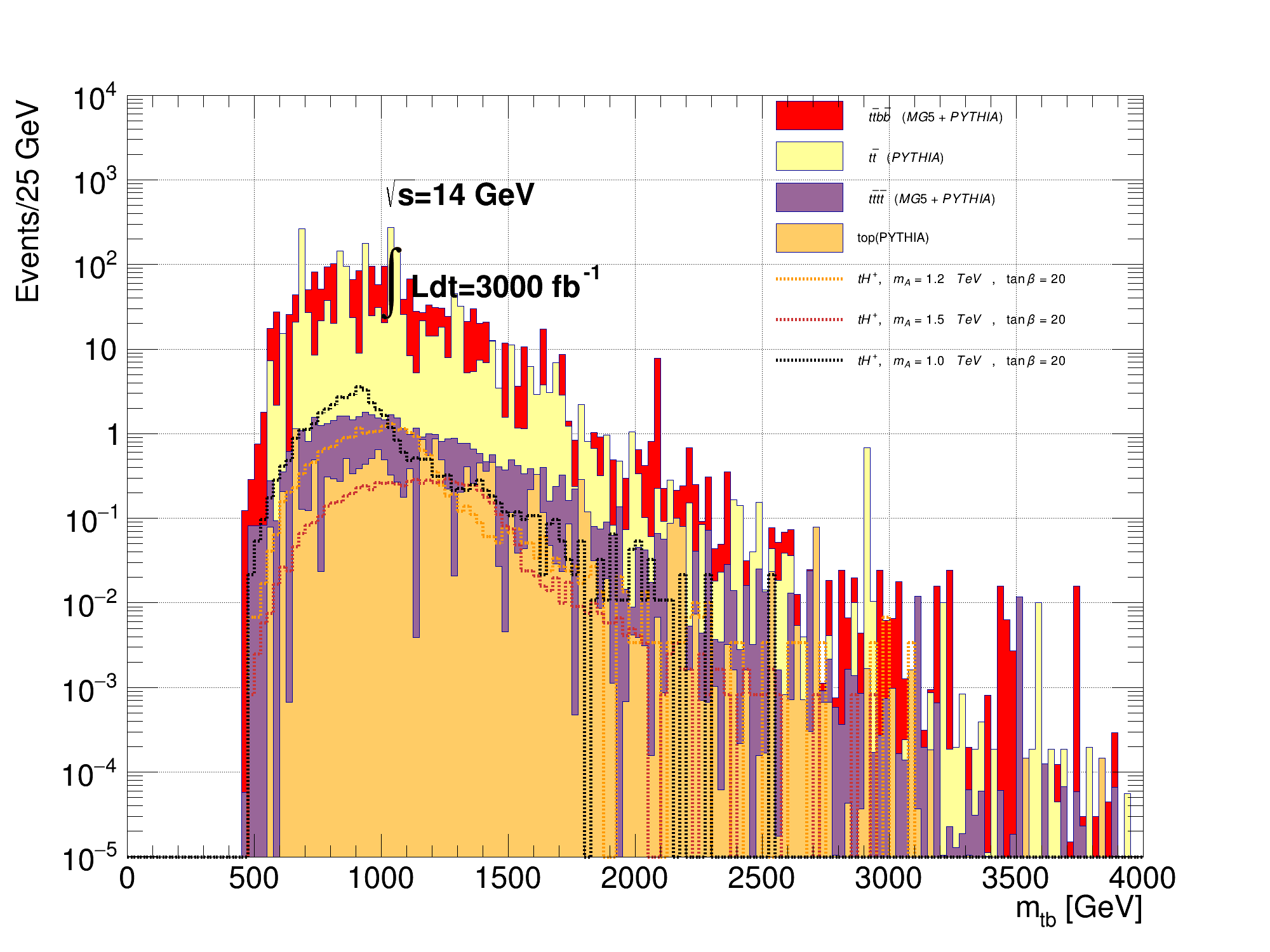}
\caption{Distribution in $m(tb)$ from $pp\to H^\pm t+X$ followed by
  $H^\pm\to tb$ decay, a fat jat tagged as a top-jet by HEPTopTagger2,
  and an isolated lepton from the decay of one of the top quarks.  We
  also show dominant SM backgrounds.
\label{fig:mtb_1t1l}}
\end{center}
\end{figure}

    The $m(tb)$ invariant mass distribution from the signal BM models
    and backgrounds, again constructed by combining $T_1$ and $b_1$, are
    shown in Fig. \ref{fig:mtb_1t1l}.  In this case, the
    $t\bar{t}b\bar{b}$ background is dominant in the range where $m(tb)$
    reconstructs $m(H^\pm )$.

\subsection{Single top (no tag) plus lepton channel}

In this channel, we examine events where the HEPTopTagger2 fails to tag
any top jets but there is a lepton from the decay of one of the tops.
We require,
\bi
\item there is at least one large radius boosted ($R < 1.5$, $p_T > 300$
  GeV) jet $n_J\ge 1$ with trimmed mass \cite{Krohn_jet_trim} $115\ {\rm
    GeV} < m_J < 190$ GeV, and at least one small radius ($R<0.4$)
  $b$-jet within the cone of the large radius jet $J$.  Then, the fat
  Jet is taken as the hardest $p_T$ top candidate directly from the
  charged Higgs decay (denoted as $J_1/T_1$).  The hardest $b$-jet
  within $J_1/T_1$ is taken as $b_2$.
\item At least 3 $b$-jets: $n_b\ge 3$, of which at least two of them
  must satisfy the signal $b$-jet requirements listed above.
\item At least 6 jets: $n_j\ge 6$.
\item Exactly one signal isolated lepton: $n_l = 1$.
\ei

We also require
\bi
\item $H_T > 1200$ GeV,
\item $m(b, b^\prime ) > 215$ GeV, where the $b$,  $b^\prime$ are the $b$
  pairs with the max $p_T$ in the events,
\item $max(R(b^\prime , H^\pm )) > 1.1$, where $b^\prime$ are any $b$-jets
  in the events that are not $b_1$ and $b_2$.
  $H^\pm$ is reconstructed from $T_1$ and $b_1$,
\item $min(R(b^\prime , b_1)) > 1$, where $b^\prime$ are any $b$-jets
  in the events and
\item $R(b_1, l) > 0.9$.
  \ei

\begin{figure}[htb!]
\begin{center}
\includegraphics[height=0.5\textheight]{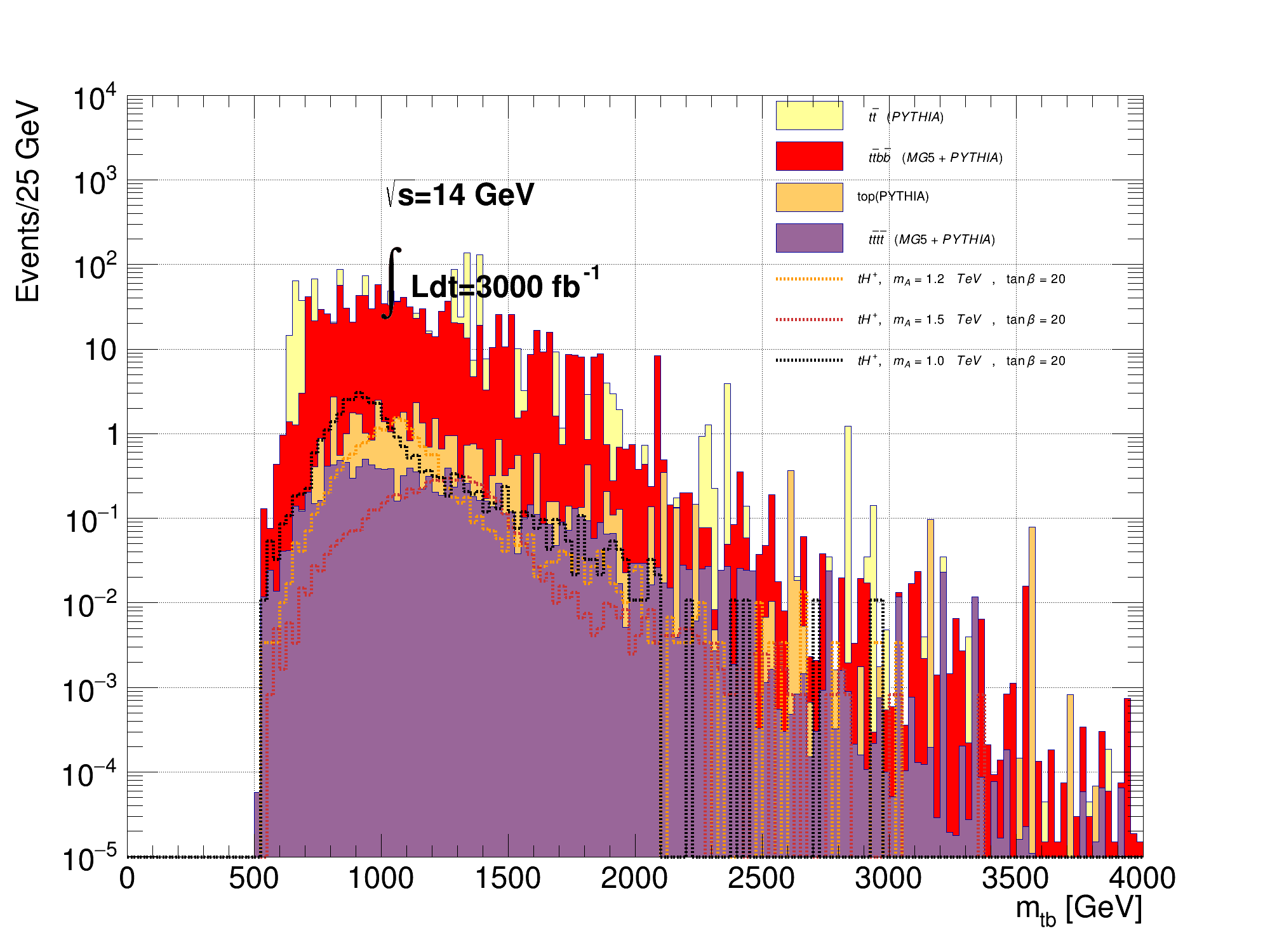}
\caption{Distribution in $m(tb)$ from $pp\to H^\pm t+X$ followed by
  $H^\pm\to tb$ decay.  We also require that there is no fat jet
  tagged as a top-jet by HEPTopTagger2 but that there is an isolated
  lepton from the decay of one of the top quarks.  We also show
  dominant SM backgrounds.
\label{fig:mtb_1t1l_notag}}
\end{center}
\end{figure}

  The $m(tb)$ distribution (again constructed by combining $T_1$ and
  $b_1$) for signal and background events with non-tagged top-jets and
  an isolated lepton is shown in Fig. \ref{fig:mtb_1t1l_notag}.

\subsection{LHC reach in $H^\pm\to tb$ channel}
\label{ssec:tb_reach}

As in the $H^\pm\to \tau\nu_\tau$ analysis, 
after adopting the above cuts for the various signal channels,
we can now create reach plots in terms of discovery sensitivity or
exclusion limits for $pp\to H^\pm t+X$ followed by $H^\pm \to tb$
in the $m_A$ vs. $\tan\beta$ plane.
In this case, we use the binned $m(tb)$ distributions
(bin width of 25 GeV) from each signal channel as displayed above
to obtain the discovery/exclusion limits.

\begin{figure}[htb!]
\begin{center}
  \includegraphics[height=0.35\textheight]{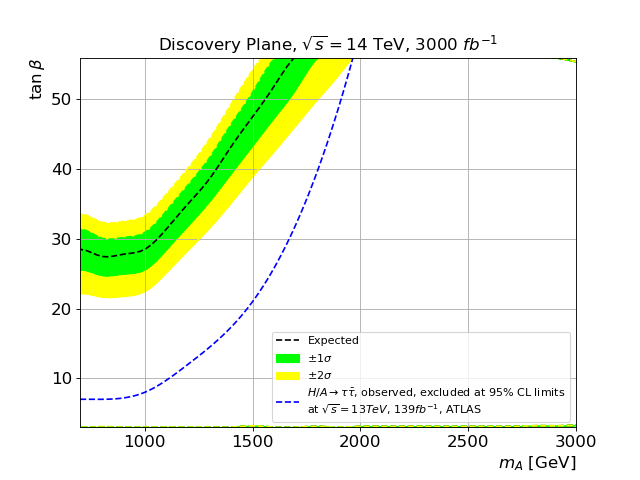}\\
  \includegraphics[height=0.35\textheight]{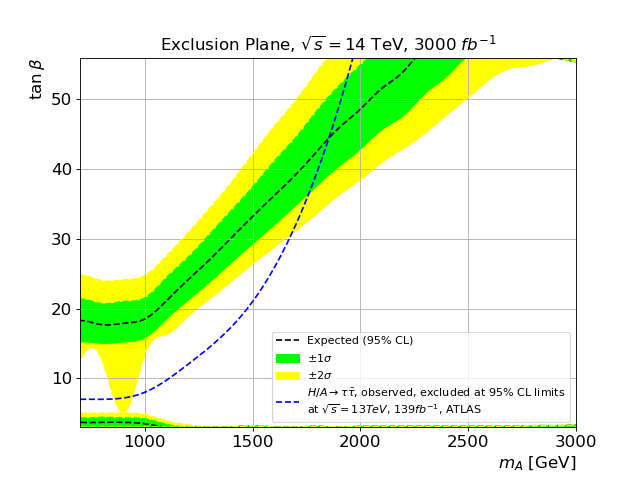}\\
  \caption{In {\it a}), we plot the $5\sigma$ discovery region of the
    $m_A$ vs. $\tan\beta$ plane for $pp\to H^\pm t+X$ followed by $H^\pm
    \to tb$ decay for HL-LHC with 3000 fb$^{-1}$.  In {\it b}), we plot
    the corresponding 95\% CL exclusion region. The dashed blue line
    shows the boundary of the region excluded at the 95\% confidence
    level in Ref.\cite{ATLAS:2020zms}.
  \label{fig:disc3000tb}}
\end{center}
\end{figure}

In Fig. \ref{fig:disc3000tb}, we show our results for the
discovery/exclusion regions via the $H^\pm\to tb$ channel for the HL-LHC
with $\sqrt{s}=14$ TeV and 3000 fb$^{-1}$ of integrated luminosity in
the $m_A$ vs. $\tan\beta$ plane using our $m_h^{125}({\rm nat})$
benchmark scenario.  In frame {\it a}), we plot the $5\sigma$ discovery
reach using the combined four $H^\pm\to tb$ signal channels listed
above.  The dashed black line denotes the computed reach while the green
and yellow bands display the $\pm 1\sigma$ and $\pm 2\sigma$
uncertainty.  From the plot, we see that a discovery region is indeed
found, starting around $m_A\sim 500$ GeV and $\tan\beta \sim 18$.  For
these combined $H^\pm\to tb$ signal channels, the discovery region
extends out to $m_A\sim 1.6$ TeV for $\tan\beta\sim 50$.  The discovery
region pinches off below $\tan\beta\sim 25$ where the 
signal, after analysis cuts, becomes too small relative to the
standard model background. Unfortunately, the entire discovery
region lies within the portion of the plane that already appears to be
excluded by the ATLAS search for $H/A \to \tau\tau$ decays
\cite{ATLAS:2020zms}.

In frame {\it b}), we plot the 95\%
CL exclusion limit for HL-LHC for our combined four signal channels.
The exclusion limit now extends out to $m_A\sim 2.1$ TeV for large
$\tan\beta\sim 50$, and well outside the ATLAS excluded region denoted
by the dashed blue line. 
We also see that the exclusion contour extends somewhat below
$\tan\beta\sim 20$ for lighter $m_{H^+}\sim 1$ TeV.
Also, a small exclusion region has now appeared at low $\tan\beta\sim 3$,
which has also appeared in the ATLAS analysis\cite{ATLAS:2021upq}.

\section{Search for $H^\pm\to$ SUSY at HL-LHC}
\label{sec:HCsusy}

In Ref. \cite{Baer:2022smj}, we examined $s$-channel production of heavy
neutral SUSY Higgs bosons followed by decays to SUSY particles, where in
natural SUSY $H,\ A\to gaugino+higgsino$ was the dominant decay mode
(when kinematically open) except where $\tan\beta$ was very large. The
heavier higgsinos decay to soft visible particles plus $\eslt$ whilst
the gauginos decay via $W,\ Z,\ h+\eslt$.  This led to discovery
channels of $H,\ A\to W,\ Z,\ h+\eslt$ at HL-LHC, with accessible
parameter regions mapped out in the $m_A$ vs. $\tan\beta$ plane for
natSUSY in Ref.\cite{Baer:2022smj}.  The number of signal events after
cuts were typically in the range of tens of events at best at HL-LHC
with 3000 fb$^{-1}$ of integrated luminosity.

The question here is then:
are there lucrative {\it charged} Higgs $H^\pm\to SUSY$ decay channels
available for HL-LHC? We saw in Fig. \ref{fig:BFHC} that for
moderate $\tan\beta$ and $m_{H^\pm}>m(gaugino)+m(higgsino)$ that the
SUSY decay modes also become the dominant decay channels for
charged Higgs bosons. And like their neutral Higgs counterparts, the
final state configurations for $H^\pm\to SUSY$ end up being
$H^\pm\to W,\ Z,\ h+\eslt$ according to the various charged Higgs and sparticle
branching fractions from Isajet\cite{Paige:2003mg}.

\begin{figure}[htb!]
\begin{center}
  \includegraphics[height=0.5\textheight]{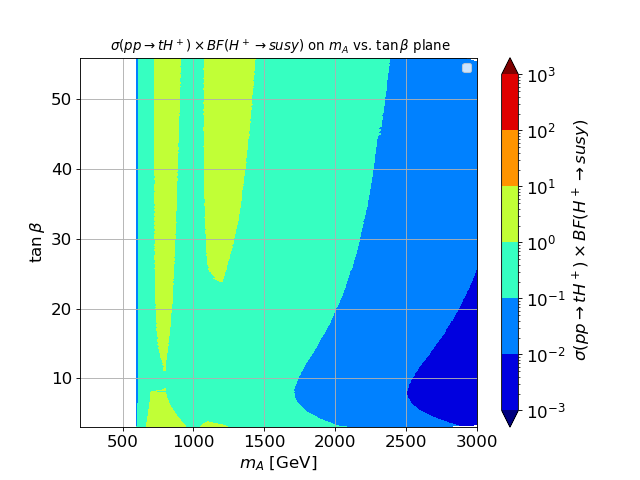}
  \caption{Plot of charged Higgs production cross section times
    $BF(H^\pm\to SUSY )$ in fb in the $m_A$ vs. $\tan\beta$ plane
    for the $m_h^{125}({\rm nat})$ scenario.\label{fig:BFsusy}}
\end{center}
\end{figure}

Thus, we have also examined the prospects for
$pp\to tH^\pm\to t+SUSY\to t+(W,\ Z,\ h)+\eslt$ at HL-LHC.
An essential difference of $pp\to tH^\pm + X$ compared to $pp\to H,\ A+X$
at LHC is that for a given heavy Higgs mass, the top-quark plus charged
Higgs cross section is typically suppressed by an order of magnitude or more
from $s$-channel neutral heavy Higgs production.
A plot of charged Higgs production cross section times
$BF(H^\pm\to SUSY )$ in fb in the $m_A$ vs. $\tan\beta$ plane
is shown in Fig. \ref{fig:BFsusy} for the $m_h^{125}({\rm nat})$ scenario.
From the plot, for $m_A\agt 1 TeV$ where decays to SUSY particles 
begin to become important, the cross sections lie in the sub-fb regime. 
In addition, one typically tries to tag the spectator $t$- or $b$- jet
in $tH^\pm$ production which also leads to a reduction in signal level.
Then one must factor in further SUSY decay branching fractions to
$W,\ Z,\ h+\eslt$ along with $W,\ Z,\ h$ branching fractions into
observable final states. The upshot is: the reduced signal channels compared
to SM backgrounds $t\bar{t}$, $t\bar{t}b\bar{b}$, $t\bar{t}W$, $t\bar{t}Z$,
$t\bar{t}h$ {\em etc.} did not lead to any compelling discovery channels that
we could find.

\section{Regions of the $m_A$ vs. $\tan\beta$ plane accessible to HL-LHC via
charged and neutral Higgs boson searches in natSUSY}
\label{sec:reach}

We have found so far that a charged Higgs boson signal should be
accessible to the HL-LHC in two different channels:
$H^\pm\to\tau\nu_{\tau}$ and, to a lesser degree, via $H^\pm\to tb$. The
regions of the $m_A$ vs. $\tan\beta$ plane which are available to HL-LHC
via $5\sigma$ discovery and $95\%$ CL exclusion have been mapped out.
At this point, it is worthwhile to compare the reach of HL-LHC via
charged Higgs boson searches to the reach which can be achieved via
$s$-channel $H$ and $A$ signals as delineated for natSUSY in Ref.
\cite{Baer:2022qqr} and \cite{Baer:2022smj}.

\begin{figure}[htb!]
\begin{center}
  \includegraphics[height=0.5\textheight]{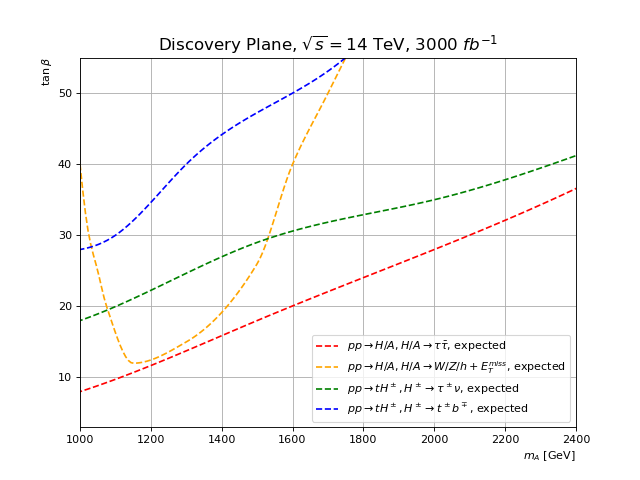}
  \caption{Plot of $5\sigma$ discovery projections for
    heavy SUSY Higgs boson searches at HL-LHC
    in the $m_A$ vs. $\tan\beta$ plane for the $m_h^{125}({\rm nat})$ scenario.}
\label{fig:disc}
\end{center}
\end{figure}

In Fig. \ref{fig:disc}, we plot out the $5\sigma$ reach of HL-LHC with
3000 fb$^{-1}$ in the $m_A$ vs. $\tan\beta$ plane for heavy charged and
neutral Higgs boson signals in the natSUSY scenario.
The red dashed contour shows the computed $5\sigma$ discovery reach
via the $H,\ A\to \tau\bar{\tau}$ channel. Of all channels assessed so far,
this provides the maximal discovery reach  due to higher production
cross sections $pp\to H,\ A +X$, lower backgrounds in the ditau decay channel
and the capability to reconstruct the ditau invariant mass $m(\tau\bar{\tau})$
using the ability to roughly reconstruct the missing neutrino momentum.
The discovery region lies above the contour whch extends from
$\tan\beta \sim 8$ for $m_A=1$ TeV to $\tan\beta\sim 35$ for $m_A=2.4$ TeV.
The decay branching fraction for $A\to \tau\bar{\tau}$
is of course enhanced by the well-known factor $\tan^2\beta$.

Next most important is the yellow-dashed contour for $pp\to H,\ A$ with
$H,\ A\to SUSY\to (W, Z,\ h)+\eslt$. These combined channels determine
the ultimate reach and only turn on for $m_A\agt 1$ TeV where $H,\ A\to
SUSY$ becomes kinematically accessible in the natSUSY scenario.  The
reach via SUSY decays is comparable with the $H,\ A\to \tau\bar{\tau}$
reach for $m_A\sim 1.2$ TeV, but for $m_A\agt 1.5$ TeV, the reach via
SUSY decays drops off more quickly for larger $\tan\beta$ values mainly
because the $H$ and $A$ decays to SM fermions are enhanced by large
Yukawa couplings, suppressing the branching fractions for decays to SUSY
particles.
The green-dashed contour denotes the HL-LHC $5\sigma$ discovery reach
via $pp\to tH^\pm+X$ followed by $H^\pm\to \tau\nu_\tau$ decay. For a
given $m_A$, the $pp\to tH^\pm+X$ cross section is well below the
resonantly enhanced $pp\to H,\ A$ and furthermore one cannot reconstruct
a charged Higgs invariant mass via the $H^\pm\to \tau\nu$ channel: as a
result, the reach via the charged Higgs channel is substantially less
than the the reach via $pp\to H,\ A\to \tau\bar{\tau}$. The blue dashed
contour denotes the $5\sigma$ discovery reach for $pp\to tH^\pm+X$ with
$H^\pm\to tb$.  This discovery region is mainly applicable at large
$\tan\beta$.  Within the model, a substantial region of parameter space
is accessible to HL-LHC via several discovery channels, but we should
keep in mind that portions of this plane is excluded by the ATLAS
search, albeit in a model with decoupled
superpartners\cite{ATLAS:2020zms}.

\begin{figure}[htb!]
\begin{center}
  \includegraphics[height=0.5\textheight]{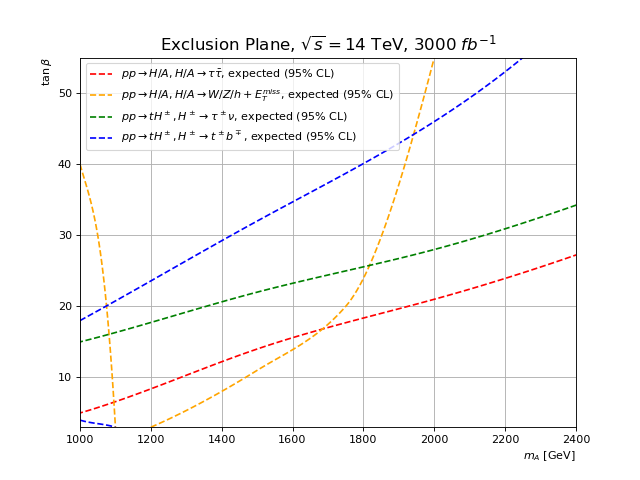}
  \caption{Plot of 95\% CL exclusion projections for
    heavy SUSY Higgs boson searches at HL-LHC
    in the $m_A$ vs. $\tan\beta$ plane for the $m_h^{125}({\rm nat})$ scenario.}
\label{fig:excl}
\end{center}
\end{figure}

In Fig. \ref{fig:excl}, we show all four contours, but now as 95\% CL
exclusion limits, should no signal appear at the HL-LHC. For $m_A\agt
1.6$ TeV, the main exclusion would come from not discovering
$H,\ A\to\tau\bar{\tau}$ while for lower $m_A\alt 1.6$ TeV the main
exclusion come from not discovering $H,\ A\to SUSY\to
(W,\ Z,\ h)+\eslt$, where the exclusion contour dips to very low
$\tan\beta\sim3$ (where $m_h$ becomes lighter than 125 GeV).  The
charged Higgs exclusion contours are contained within the 
$H,\ A\to\tau\bar{\tau}$ exclusion contour.

\section{Conclusions}
\label{sec:conclude}

In this paper we have investigated the ability of HL-LHC to discover the
charged Higgs bosons of supersymmetric theories in the natural SUSY scenario.
We believe that the natSUSY scenario is strongly motivated in that it naturally
explains the measured magnitude of the weak scale which arises from a conspiracy
of the weak scale soft terms: $\mu\sim 100-400$ GeV while $m_{H_u}^2$ is
radiatively driven to rather small negative values at the weak scale
(EW symmetry is barely broken). Other sparticle masses can be much larger,
lying in the TeV or beyond region since their contributions to the weak scale
are suppressed (at least) by loop factors. Unnatural SUSY models which predict
much higher values for $m_{weak}$ are regarded as rather implausible since they
will require an unnatural conspiracy/finetuning of parameters in order to
gain $m_{weak}\sim 100$ GeV.

The charged Higgs boson masses can range from their present lower limits
from LHC searches up into the multi-TeV range (depending on $\tan\beta$)
with little cost to EW naturalness. Furthermore, once their masses
exceed $m(higgsino)+m(gaugino)$, then the decay to SUSY particles can
become dominant.  This reduces heavy Higgs decay to SM particle signals
as expected in unnatural scenarios such as 2HDMs, but also opens up
possible new avenues for heavy Higgs discovery. In this paper, we have
delineated search strategies for charged Higgs bosons in both the
$H^\pm\to\tau\nu_{\tau}$ and $H^\pm\to tb$ channels and have also
computed the regions of $m_A$ vs. $\tan\beta$ parameter space which are
accessible to HL-LHC as the $5\sigma$ and 95\% CL contours.  The HL-LHC
reach for charged Higgs bosons is typically contained within the
stronger reach via $s$-channel production of heavy neutrals $H$ and $A$.
For searches at the HL-LHC, decays of the charged Higgs boson to
superpartners appear to be unimportant.  Nonetheless, there do exist
regions where all of $H^\pm$, $H$ and $A$ may be discovered.

{\it Acknowledgements:} 

This material is based upon work supported by the U.S. Department of Energy, 
Office of Science, Office of High Energy Physics under Award Number DE-SC-0009956 and DE-SC-0017647.


\bibliography{HC}
\bibliographystyle{elsarticle-num}

\end{document}